# Direct Implicit and Explicit Energy-Conserving Particle-in-Cell Methods for Modeling of Capacitively-Coupled Plasma Devices

Haomin Sun[1*], Soham Banerjee[2], Sarveshwar Sharma[3,4*], Andrew Tasman Powis[5], Alexander V. Khrabrov[5], Dmytro Sydorenko[6], Jian Chen[7], Igor D. Kaganovich[5]

[1] École Polytechnique Fédérale de Lausanne (EPFL), Swiss Plasma Center (SPC), CH-1015 Lausanne, Switzerland

[2] Birla Institute of Technology and Science, Pilani 333031, India

[3] Institute for Plasma Research, Bhat, Gandhinagar 382428, India

[4] Homi Bhabha National Institute, Anushaktinagar, Mumbai, Maharashtra 400094, India

[5] Princeton Plasma Physics Laboratory, Princeton University, Princeton, New Jersey 08543, USA

[6] University of Alberta, Edmonton, Alberta T6G 2E1, Canada

[7] Sino-French Institute of Nuclear Engineering and Technology, Sun Yat-sen University, Zhuhai 519082, P. R. China

Corresponding Authors: Haomin Sun, Email: haomin.sun@epfl.ch

Sarveshwar Sharma, Email: sarvesh@ipr.res.in, sarvsarvesh@gmail.com

Achieving large-scale kinetic modelling is a crucial task for the development and optimization of modern plasma devices. With the trend of decreasing pressure in applications such as plasma etching, kinetic simulations are necessary to self-consistently capture the particle dynamics. The standard, explicit, electrostatic, momentum-conserving Particle-In-Cell method suffers from restrictive stability constraints on spatial cell size and temporal time step, requiring resolution of the electron Debye length and electron plasma period respectively. This results in a very high computational cost, making the technique prohibitive for large volume device modeling. We investigate the Direct Implicit algorithm and the explicit Energy Conserving algorithm as alternatives to the standard approach, both of which can reduce computational cost with a minimal (or controllable) impact on results. These algorithms are implemented into the well-tested EDIPIC-2D and LTP-PIC codes, and their performance is evaluated via 2D capacitively coupled plasma discharge simulations. The investigation revels that both approaches enable the utilization of cell sizes larger than the Debye length, resulting in reduced runtime, while incurring only minor inaccuracies in plasma parameters. The Direct Implicit method also allows for time steps larger than the electron plasma period, however care must be taken to avoid numerical heating or cooling. It is demonstrated that by appropriately adjusting the ratio of cell size to time step, it is possible to mitigate this effect to an acceptable level.

## I. Introduction

Simulation and modeling of plasma devices plays an important role in the development and optimization of modern plasma reactors due to the high cost of experimentation over a wide range of operating conditions[1, 2]. Traditional methods for modelling plasma devices include the fluid approach[3-10], hybrid fluid/kinetic models[11] and kinetic methods such as the Particle-in-Cell (PIC) technique[12, 13]. In the fluid approach, fluid equations are solved for all plasma species under the assumption that the particle velocity distribution function (VDF) is Maxwellian. The greatest advantage of this approach is a smaller computational cost comparing with kinetic simulations[8, 14-16], nominally because the assumed VDF reduces the number of equations to solve. However, important kinetic effects are neglected in this approach, when the VDF cannot be the presumed, and therefore the accuracy of such a model is questionable in industrial processing devices operating at low pressures (where the mean free path is on the order of or larger than the sheath size)[17-19]. PIC methods, are often used to study the kinetic effects by solving the Vlasov equation[20-30], with the most common method being the traditional explicit momentum-conserving algorithm. For the electrostatic explicit momentum-conserving algorithm, the time step is limited by the fastest electron dynamics ($\omega_{pe} dt \leq 0.2$), while the cell size is limited by the Debye radius ($dx \leq \lambda_{De}$). Here, $\omega_{pe}(= \sqrt{n_0 e^2 / \varepsilon_0 m_e})$ and $\lambda_{De}(= \sqrt{\varepsilon_0 T_e / n_0 e})$ are the electron plasma frequency and Debye radius; $dt$ and $dx$ are the time step and cell size, respectively. The requirement for such a fine resolution results in very high computational cost.

The current trend in plasma processing is to operate at lower gas pressures ($\sim$ several mTorr) [31-35] to achieve atomic-scale precision. The capacitively coupled plasma (CCP) reactor, which operates in the MHz range of radio-frequencies (RF)[36, 37], is the most widely used contemporary plasma device for etching of silicon wafers, critical to the semiconductor industry. There have been a large number of modeling and experimental work exploring various aspects of CCP discharges. In single-frequency capacitively coupled plasma discharges, the ion flux and ion energy cannot be controlled independently since they depend simultaneously the input power for a fixed neutral gas pressure and system geometry[32, 38-47]. Several other approaches have been suggested to circumvent this restriction, such as Dual-Frequency Capacitively Coupled Plasma (DF-CCP) discharges[48-55], electrical asymmetric effects[56-60], and non-sinusoidal tailored voltage/current waveform excitations[61-69]. There are also several fascinating physical phenomena in Very High Frequency (VHF) driven CCP discharges, such as the generation of an electron beam accelerated by the sheath electric fields[70-72] as well as the production of higher harmonics in voltage and current[73-80].

These devices require modeling of plasma chemistry with numerous plasma and radical species, meaning that even 1D-3V PIC simulations can become computationally expensive.

Modern plasma devices, on the other hand, often operate with plasma density up to the order of $10^{17} m^{-3}$, and the size of a typical CCP discharge chamber is at least $\sim 10\ cm$. For a realistic 2D simulations of a plasma discharge, this requires $\sim 10^6$ cells and usually $\sim 10^7$ time steps to achieve convergence. Furthermore, in order to reduce numerical noise, the number of macro-particles in each cell should be higher than $100$[81, 82], requiring hundreds of millions or billions of simulation particles evolved over 10s of millions of time steps. To achieve reasonable simulation time, one faces a difficult dilemma, whether to reduce the simulation size of the modeled system, which could alter some plasma properties, or spend significant computational time, often weeks or sometimes even months, to achieve a steady state solution. Therefore, it is essential to remove the grid size and time step restrictions on the PIC method if we are to model the whole plasma device and reduce the actual simulation run time, so that these new CCP reactors can be modeled for prototyping.

The Direct Implicit Particle-In-Cell (DI-PIC) method, was developed in 1980s to provide such a reduction in computational cost, with the added benefit of stability even when under-resolving the Debye length and plasma frequency[83-86], allowing for a larger time step and cell size. Within a CCP discharge resolution of the fast plasma electron dynamics with time scale $t \sim \omega_{pe}^{-1}$ is not crucial, meaning that accuracy can be maintained even when the electron time and length scales are under-resolved.

Similarly, the explicit Energy Conserving Particle-In-Cell (EC-PIC) algorithm proposed by Lewis[87] alleviates the constraint on the cell size ($dx$) by making small modifications to the traditional PIC method. Despite several efforts to study the energy conserving and direct implicit methods, most studies have focused on uniform plasmas (using periodic boundary conditions). Other early DI-PIC modelling efforts were limited to 1D simulations of RF discharges[88, 89]. Applying the DI-PIC approach in two-dimensions requires a more complex field solver and particle pusher. Other advanced implicit algorithms[90-93] have also been proposed. However, these approaches are more computationally expensive per time step, and are complex to implement when compared to the much simpler DI-PIC and EC-PIC studied in this paper.

DI-PIC and EC-PIC are implemented into the EDIPIC-2D code[94] and EC-PIC is implemented into the LTP-PIC code[95] (see a full description in Section 3.4) and their accuracy and performance are tested through simulations of a 2D capacitively coupled plasma discharge. It is found that both approaches allow for cell sizes larger than the Debye length, and a corresponding reduction in runtime, with only a small penalty in accuracy. The Direct Implicit method can also allow for time steps larger than the electron plasma period, however this can lead to numerical heating or cooling. In these cases, the tuning of the cell size to time step ratio can reduce this heating/cooling to tolerable levels.

To model the collision operator, we use the well-known null-collision algorithm for Monte-Carlo Collisions (MCC), and approach which has been proven to be accurate and efficient at reducing numerical cost when modeling many applications of low-pressure devices [22, 96, 97].

We performed a large number of CCP simulations with both the DI and EC methods, as well as benchmark them against results using explicit momentum conserving PIC simulations. We determined that the computational efficiency is significantly improved while still maintaining sufficient accuracy. For DI-PIC, the accuracy can be further improved if the initial time step and cell size are chosen according to an "optimum path" recipe to be described further in the paper. An estimate of numerical heating in the direct implicit method is also provided, which enables trustworthy modelling of large plasma devices.

The paper is organized as follows: The simulation algorithm and model are introduced in Section 2. Code tests and benchmarks are presented in Section 3. Finally, the discussion of results and conclusions are given in Section 4.

## II. Numerical Algorithm and Simulation Model
### II.1 Direct Implicit (DI) PIC method

The Direct Implicit (DI) Particle-in-Cell method predicts the charge density at the next time step using a linear approximation of the new electric field[98]. This approach yields an elliptic equation for the potential whose coefficients depend directly on particle data accumulated on the spatial grid in the form of an effective linear susceptibility. The rank of the matrix equation is determined by the number of field quantities defined on the spatial grids, so that potential profile can be solved using standard numerical linear algebra techniques. The particle update is calculated for each particle concurrently[84, 86]. The DI time advance loop is comprised of a predictive push for the particles, followed by solving the electrostatic field equation and then by a second, final particle push. In the presence of an external magnetic field, the electrostatic field equation takes the form of a Poisson equation for an anisotropic dielectric. On a Cartesian grid, the corresponding finite-difference approximation involves a nine-point "box" type stencil, compared to the five-point "star" appearing in the explicit scheme[99]. Also, since the effective dielectric tensor in the DI scheme is a function of the local charge density, the matrix supplied to the field solver must be re-initialized at each time step. In the un-magnetized case, the box stencil reduces to the five-point "star" stencil. A flow chart for the DI-PIC method is shown in Fig. 1.

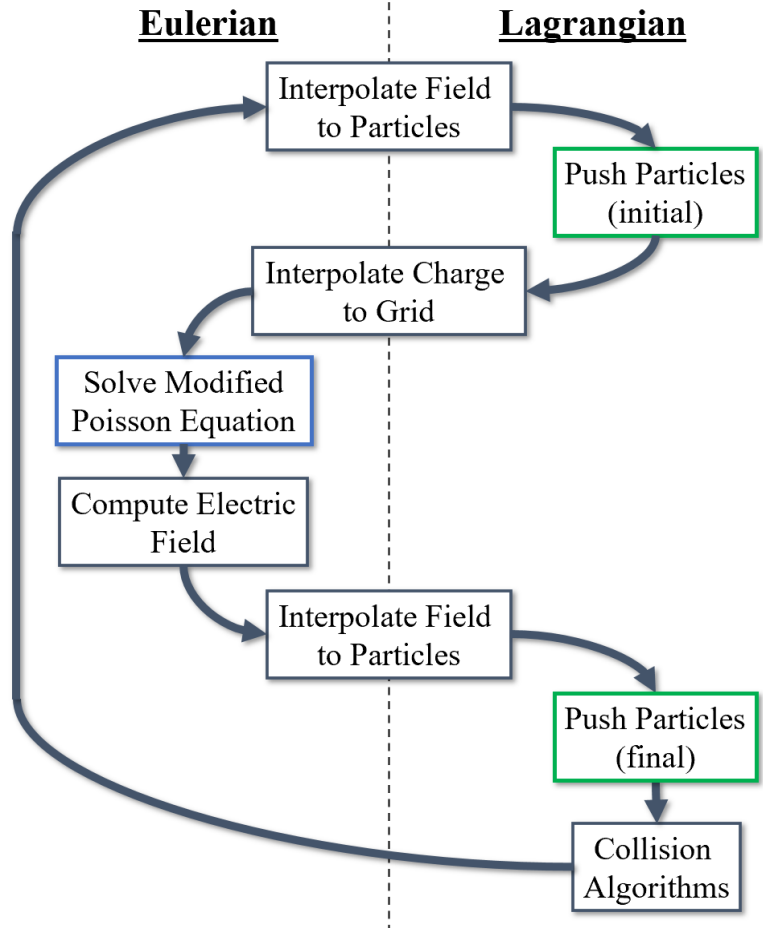

**Eulerian** — **Lagrangian**

Interpolate Field to Particles

Push Particles (initial)

Interpolate Charge to Grid

Solve Modified Poisson Equation

Compute Electric Field

Interpolate Field to Particles

Push Particles (final)

Collision Algorithms

**Figure 1**: Flowchart for the Direct Implicit PIC algorithm. Compared to the traditional Momentum Conserving algorithm the particles undergo two updates (shown in green boxes) as per Eqs. (5)-(6) and (12)-(14), the Poisson equation (shown in blue box) is also modified as per Section 2.1.2. Note that in EDIPIC interpolation to/from the grid occurs within the same kernel as the particle update.

We begin with a generalized case, a 2D plasma with external magnetic field $\boldsymbol{B} = \boldsymbol{B}(x, y)$. We initially introduce the following notations:

$$\boldsymbol{\Omega} = \frac{q\boldsymbol{B}}{mc}, \tag{1}$$

$$\theta = \frac{1}{2}\boldsymbol{\Omega}dt, \tag{2}$$

$$(A^{-1})_{\alpha\beta} = \frac{1}{1+\theta^2}(\delta_{\alpha\beta} + \theta_\alpha\theta_\beta + e_{\alpha\beta\gamma}\theta_\gamma), \tag{3}$$

$$\epsilon_{\alpha\beta} = \delta_{\alpha\beta} + \frac{1}{2}(\omega_{pe}\Delta t)^2 (A^{-1})_{\alpha\beta}. \tag{4}$$

Note that $q = -e$ in the case of electron species to which the DI treatment is being applied. In the expressions above, $\boldsymbol{\Omega}$ is the particle cyclotron frequency, $\theta$ is a dimensionless notion of $\boldsymbol{\Omega}$ in the algorithm, $A^{-1}$ and $\epsilon_{\alpha\beta}$ are defined as the field matrix, $\delta_{\alpha\beta}$ is the unitary tensor, $e_{\alpha\beta\gamma}$ is the anti-symmetric tensor, and $\alpha, \beta, \gamma$ denote different spatial dimensions.

### 2.1.1 Particle advance: predictive push

Making use of the notations introduced above, the predictive time-advance step for electrons is carried out as follows:

$$\tilde{\boldsymbol{v}} = K^n \boldsymbol{v}^{n-1/2} + dt(A^{-1})^n \left( \frac{1}{2} \boldsymbol{a}^n + \frac{q}{m} \boldsymbol{E}^n_{ext} \right), \qquad (5)$$

$$\tilde{\boldsymbol{x}} = \boldsymbol{x}^n + \tilde{\boldsymbol{v}} dt, \qquad (6)$$

where we have:

$$\boldsymbol{K} = \frac{1}{2}(\boldsymbol{I} + \boldsymbol{A^{-1}}), \qquad (7)$$

and the matrix $\boldsymbol{A^{-1}}$ is given by Eq. (3). Here, $\tilde{\boldsymbol{v}}$ and $\tilde{\boldsymbol{x}}$ are the particle velocity and position after the predictive push, $\boldsymbol{v}^{n-1/2}$ and $\boldsymbol{x}^n$ are the particle velocity and position before the predictive push, $\boldsymbol{E}^n_{ext}$ is the external interpolated electric field from the grids to the particle positions, $\boldsymbol{I}$ is the unitary matrix, and $m$ is particle mass. The particle phase variables updated at the predictive step are also known as the "streaming" values. It is seen that an additional quantity, the acceleration vector $\boldsymbol{a}$, has been added to the data structure for electron particles. Corresponding modifications have been made to the explicit EDIPIC-2D code[94] to handle particle transfer between grid blocks used for parallelization, as well as for load balancing (now taking into account the different lengths between the data structures carried by electrons and ions)[94]. The grid-accumulated "streaming" electron charge density $\tilde{\rho}_e$ collected after the predictive push is supplied to the field solver. The procedure for collecting the ion contribution to the charge density is the same as for electrons.

### 2.1.2 Field Solver

The field equation arising in the DI scheme are:

$$\nabla \cdot \boldsymbol{D}^{n+1} = -\frac{(\tilde{\rho}_e + \rho_i)}{\epsilon_0}, \tag{8}$$

Where,

$$D_\alpha = \epsilon_{\alpha\beta} \partial_\beta \Phi, \tag{9}$$

is the electric displacement field vector and the $\epsilon$ tensor is given by Eq. (4). The grid-accumulated electron charge density $\tilde{\rho}_e$ is used to evaluate the components of $\epsilon$. Here $\rho_i$ and $\Phi$ are the ion charge density and potential, respectively. The grid values of the matrix $\boldsymbol{A^{-1}}$ evaluated at half-nodes ("shifted" values), which is required to evaluate $\epsilon$, are tabulated prior to executing the main loop. Equation (8) is solved for the grid values of the potential $\Phi$ using the PETSc library[100].

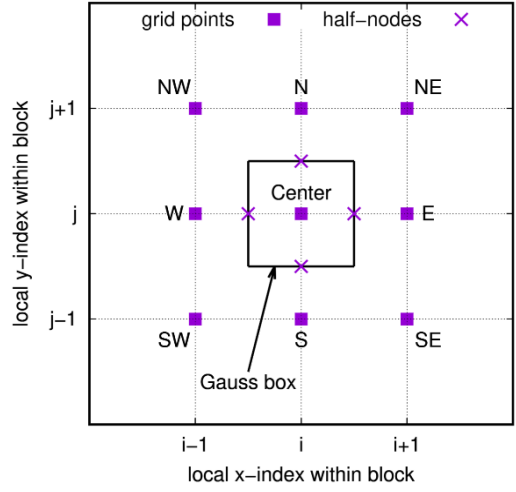

**Figure 2**: The 9-point stencil and the Gauss integration box for the discretized direct implicit Poisson equation are shown here. The components of the dielectric tensor are interpolated from the grid to half-nodes. The electric field at half nodes is approximated by differentiating the potential.

Equation (8) is discretized with the help of Gauss's law with the finite-difference stencil shown in Fig. 2. It is convenient to identify the coefficients of the linear equation arising in the differencing scheme according to the points of the compass. The four-unit normal vectors involved in evaluating the flux of the $\boldsymbol{D}$ vectors are pointing North (N), South (S), East €, and West (W). The Northwest (NW), Northeast (NE), Southeast (SE) and Southwest (SW) directions are also shown in the figure. The $x$ and $y$ components of the electric field **E** are calculated at the half-nodes through the second order central differencing scheme.

Let us introduce the values of $\epsilon$ components at the half nodes adjacent to the $(i, j)$ node where the charge density is specified as $\epsilon^E$, $\epsilon^W$, $\epsilon^N$, and $\epsilon^S$. We also introduce the following auxiliary quantities:

$$a_{NS} = \frac{1}{4}(\epsilon_{xy}^W - \epsilon_{xy}^E), \tag{10}$$

and

$$a_{EW} = \frac{1}{4}(\epsilon_{xy}^N - \epsilon_{xy}^S). \tag{11}$$

Then, the matrix coefficients contributed by the given locally numbered $(i, j)$ node where the charge density is specified on the right-hand side are given in Table 1.

**Table 1**: Elements of the matrix for the field solver in direct implicit algorithm.

| Node | Coefficient |
|------|-------------|
| **South** | $-a_{NS} - \epsilon_{yy}^S$ |
| **North** | $a_{NS} - \epsilon_{yy}^N$ |
| **West** | $a_{EW} - \epsilon_{xx}^W$ |
| **East** | $-a_{EW} - \epsilon_{xx}^E$ |
| **NE** | $-0.25(\epsilon_{xy}^E + \epsilon_{yx}^N)$ |
| **SE** | $0.25(\epsilon_{xy}^E + \epsilon_{yx}^S)$ |
| **SW** | $-0.25(\epsilon_{xy}^W + \epsilon_{yx}^S)$ |
| **NW** | $0.25(\epsilon_{yx}^N + \epsilon_{xy}^W)$ |
| **Center** | $\epsilon_{xx}^E + \epsilon_{xx}^W + \epsilon_{yy}^S + \epsilon_{yy}^N$ |

Modifications were made to calculate the elements of the matrix contributed by the nodes located at the material boundaries. Within a dielectric object with relative permittivity $\epsilon_d$ the dielectric tensor (see Eq. (8)-(9)) takes the form:

$$\epsilon_{\alpha\beta} = \epsilon_d \delta_{\alpha\beta}.$$

The resulting properties of the DI permittivity tensor were taken into account to calculate the respective elements of the matrix fed to the field solver.

### 2.1.3 Particle advance: final push

The electric field calculated by the field solver, $E^{n+1} = -\nabla\Phi^{n+1}$ on the staggered grids, is interpolated to the electron particle position obtained at the predictive push, and the final push is performed as follows:

$$\delta v = \frac{q\Delta t}{2n} K^n E^{n+1}(\widetilde{x}), \tag{12}$$

$$v^{n+1/2} = \widetilde{v} + \delta v, \tag{13}$$

$$a^{n+1} = \frac{1}{2}\left(a^n + \frac{q}{m} E^{n+1}\right). \tag{14}$$

Note that in the DI scheme, the $K$ matrix for the final push is evaluated at the particle position corresponding to the previous time step[84]. In order to evaluate $K^n$, the electron particle positions are back-tracked in order to reduce memory consumption.

### 2.1.4 Quantify numerical heating or cooling

One of the important properties that limits the performance of the DI algorithm is the presence of numerical heating or cooling. This arises since the DI algorithm does not guarantee exact energy conservation, which is due to the dissipative nature of the implicit time integration scheme[84, 85]. However, previous works has demonstrated that these effects can nearly vanish when the time step ($dt$) and cell size ($dx$) follow an "optimal path", which is a linear relation between time step and cell size[83-85]. Therefore, the numerical heating/cooling of the DI algorithm is controllable provided we choose the time step and cell size appropriately. Here, we verify this important property using our in-house 2D-3V code by simulating a double periodic system with initial plasma density $n_{e0} = 1.1 \times 10^{14}\ m^{-3}$ and initial temperature $T_{e0} = 1\ eV$. A double periodic system typically refers to a 2D system where all the four boundaries are periodic, where particles moving out from one boundary are added to the other side with the same velocity. This system also allows for the Poisson equation to be solved via Fourier techniques. In each cell, we have $N = 100$ macro-particles. We alter the cell size and time step and compare how the numerical heating/cooling affects the temperature of the electrons. The normalized numerical heating $\Delta E/E_0 N$ for different cases at $t = 591.2\omega_{pe}^{-1}$ are shown in Table 2.

**Table 2**: Normalized numerical heating $(\Delta E/E_0)/N$ at $t = 591.2\omega_{pe}^{-1}$ for cases with different time step and cell size. The optimal cases are denoted in red.

| $\left(\dfrac{\Delta E}{E_0}\right)/N$ | | $\omega_{pe}dt$ | | | | | |
|---|---|---|---|---|---|---|---|
| | | 0.2 | 0.4 | 0.8 | 1.6 | 3.2 | 6.4 |
| | 32 | | | | $1.3 \times 10^{-4}$ | $-4.8 \times 10^{-6}$ | $-2.2 \times 10^{-4}$ |
| | 16 | $2.7 \times 10^{-4}$ | $2.2 \times 10^{-4}$ | $1.2 \times 10^{-4}$ | $-7 \times 10^{-6}$ | $-1.8 \times 10^{-4}$ | |
| $dx/\lambda_{De}$ | 8 | $5 \times 10^{-5}$ | $5.7 \times 10^{-5}$ | $-2 \times 10^{-7}$ | $-1.3 \times 10^{-4}$ | | |
| | 4 | $1.5 \times 10^{-5}$ | $4.8 \times 10^{-6}$ | $-6.8 \times 10^{-5}$ | | | |
| | 2 | $-3.3 \times 10^{-7}$ | $-2 \times 10^{-5}$ | | | | |
| | 1 | $-3.2 \times 10^{-6}$ | | | | | |

A near-zero numerical heating is possible when the ratio between $\omega_{pe}dt$ and $dx/\lambda_{De}$ is approximately 0.1. We refer to this as the "optimal path" of the DI algorithm, since using these time steps and cell sizes provides near energy-conservation, which is beneficial for long time-scale modelling of plasma devices.

The numerical heating can be quantified by linearly fitting the data points:

$$y = ax + b$$

where $y = \Delta E/E_0 N$, $x = v_{th,e}dt/dx$, $a = 5.11 \times 10^{-6}$, $b = -7.80 \times 10^{-5}$, with fitting accuracy $R^2 = 0.8238$. Extending this analysis to non-periodic systems will be discussed in Sec. 3.3.

## 2.2 Energy Conserving (EC) PIC method

The explicit Energy Conserving PIC method (EC-PIC) here refers to the scheme originally reported by Lewis in 1970[87], whereby a variation approach is adopted for deriving the appropriate interpolation functions and Poisson equation/electric field finite difference discretizations so as to conserve system energy in the limit $dt \to 0$. In practice, energy remains very well-conserved when adhering to the standard MC-PIC time constraint ($\omega_{pe}dt < 0.2$), making the scheme resilient to the finite-grid-instability, even for $dx \gg \lambda_{De}$ [87]. EC-PIC schemes for both electrostatic and electromagnetic PIC codes have been derived in References[101, 102] and placed on a more sound theoretical footing through the language of discrete exterior calculus[103]. Initial investigations into the scheme revealed a deleterious cold-beam, finite-grid instability which quickly resulted in unphysical behavior[104]. However, Ref. [99] demonstrated that explicit EC-PIC could safely be applied to warm plasmas primarily driven by ambipolar electric fields, such as the system investigated in this paper. Therefore, the EC-PIC method could prove to be a useful method to model large scale plasma systems without the need to resolve the Debye length, an issue this paper will aim to explore.

The EC-PIC scheme is very similar to the traditional MC-PIC approach with a few important exceptions. A flow chart for EC-PIC scheme is shown in Fig. 3.

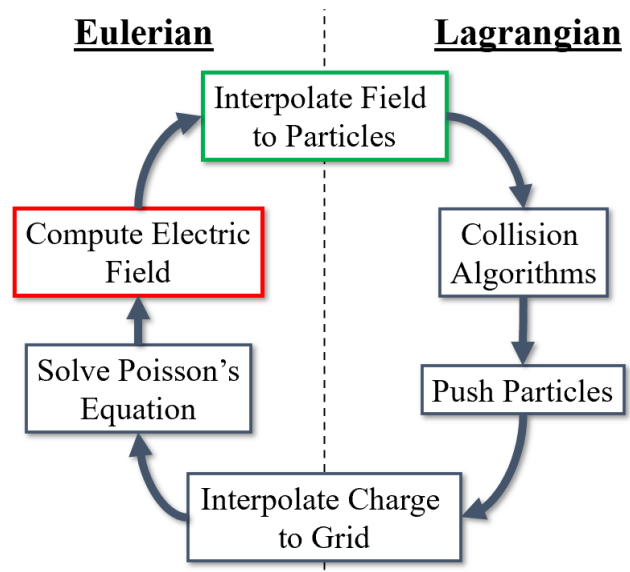

**Figure 3**: Flowchart for the explicit Energy Conserving PIC algorithm. The algorithm structure is identical to the standard Momentum Conserving algorithm, however the interpolation of electric field to the particles (shown in green box) is modified as per Eqs. (17)-(20) and calculation of the electric field from potential is modified as per Eqs. (15)-(16) (shown in red box).

The algorithm proceeds as follows:

1. Particles are updated via the standard Boris algorithm [102].

2. Particle charge density is interpolated to the grid via multi-linear (in this instance bilinear) or cloud-in-cell interpolation.

3. The Poisson equation in EC-PIC is solved using exactly the same method as in explicit MC-PIC code, where the standard a five-point stencil is implemented.

4. Modification: The electric field is computed at the edge centers between cell nodes via the central difference scheme. For example, consider discrete electric potential $\phi_{i,j}$ on a 2D lattice with $i$ and $j$ representing the discrete $x$ and $y$ coordinates, respectively. The electric field components are computed as:

$$E_{x,i+1/2,j} = -\frac{\phi_{i+1,j}-\phi_{i,j}}{dx}, \tag{15}$$

$$E_{y,i,j+1/2} = -\frac{\phi_{i,j+1}-\phi_{i,j}}{dy}. \tag{16}$$

5. Modification: Interpolation of the electric field to the particles is performed using $0^{\text{th}}$ order (nearest grid point) in the same direction as the electric field component and $1^{\text{st}}$ order (linear) in all other directions. For a particle $p$ located within cell $i,j$ with node $i,j$ at the South-West corner, the electric field components are defined as:

$$E_{x,p} = \sum_{i,j} E_{x,i+1/2,j} S_0\big(x_p - x_i + dx/2\big) S_1\big(y_p - y_j\big) \tag{17}$$

$$E_{y,p} = \sum_{i,j} E_{y,i,j+1/2} S_1\big(x_p - x_i\big) S_0\big(y_p - y_j + dy/2\big) \tag{18}$$

where

$$S_0(x) = \begin{cases} 1, & -dx/2 \leq x \leq dx/2 \\ 0, & otherwise \end{cases} \tag{19}$$

$$S_1(x) = \begin{cases} 1 + x/dx, & -dx \leq x \leq 0 \\ 1 - x/dx, & 0 \leq x \leq dx \end{cases} \tag{20}$$

Here $S_0$ and $S_1$ are the zero order and first order particle shape functions. Note that other shape functions can also be used to provide higher order interpolation[92].

## 3. Comparison of the Numerical Methods

### 3.1 Simulation Setup

To perform code testing and benchmarking, we model a two-dimensional (2D) Capacitively Coupled (CCP) discharge using the traditional explicit Momentum Conserving PIC (MC-PIC), and our newly developed Direct Implicit algorithm (DI-PIC) and explicit Energy Conserving PIC (EC-PIC) in EDIPIC-2D. The EC-PIC results are further benchmarked our in-house Low-Temperature Plasma Particle-in-Cell (LTP-PIC) code. A schematic diagram of the simulation domain is shown in Fig. 4, with $L_x \times L_y = 108\,mm \times 36\,mm$. A powered electrode with voltage amplitude $V_{rf} = 100\,V$ and $f_{rf} = 27.12\,MHz$ is placed at the bottom edge, while the other three boundaries are grounded electrodes. A small vacuum gap of 2.25 mm is implemented between the powered electrode and the grounded electrodes. The initial plasma temperature and density are $T_{e0} = 2\,eV$ and $n_{e0} = 5 \times 10^{15}\,m^{-3}$, respectively. Argon is used as the working neutral gas, with pressure $P_g = 5\,mTorr$ for all simulations. We include electron-neutral elastic, excitation (at 11.15 eV only) and ionization collisions as well as ion-neutral charge exchange. Tables for electron-neutral cross-sections are from Ref. [105, 106] and the semi-empirical model from Ref. [107] is used to compute ion charge-exchange cross-sections. We initialize all simulations with 100 macro-particles per-cell for all cases and run for $102\,\mu s$, corresponding to more than 2,500 rf cycles, at which point steady state is reached. We performed two simulations using MC-PIC with $dx_{explicit} = 7.5 \times 10^{-5}\,m \approx \lambda_{De,0}/2$, $dt_{explicit} = 10^{-11}\,s$ as wall as with $dx_{explicit} = 1.5 \times 10^{-4}\,m \approx \lambda_{De,0}$, $dt_{explicit} = 10^{-11}\,s$ to ensure numerical convergence. We use the most highly resolved of these cases as a reference to illustrate the accuracy of explicit EC-PIC and implicit DI-PIC methods (this reference case is referred to as "explicit" in figure captions), when increasing systematically both the time step and cell size, such that $dx$ may be greater than the Debye length $\lambda_{De}$ and time step larger than the electron plasma frequency, i.e. $\omega_{pe}^{-1}$. All simulation parameters are shown in Table 3, where we set $dx_0 = \lambda_{De,0}$, $dt_0 = dt_{explicit}$ as the cell size and the time step, respectively.

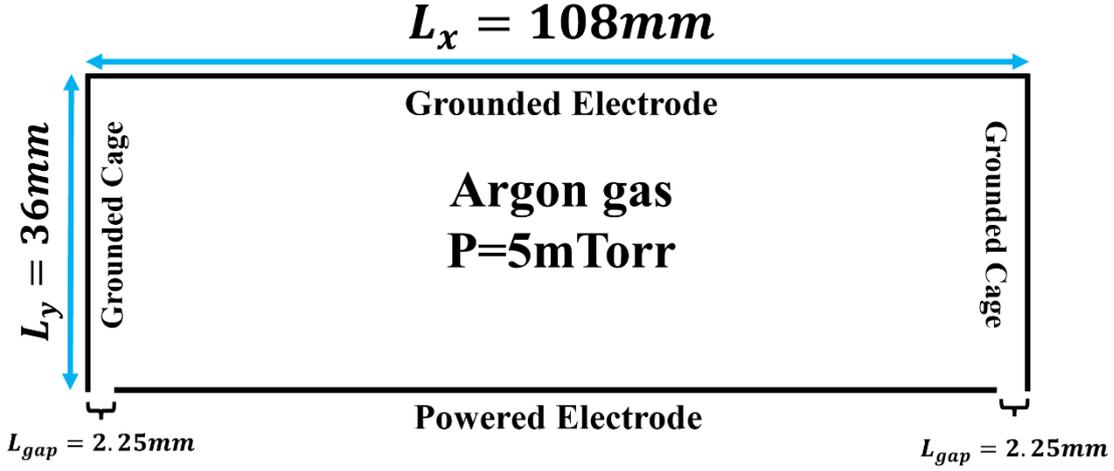

**Figure 4**: Schematic diagram of the Argon CCP simulation domain.

**Table 3**: Simulation cases studied with different cell sizes and time steps.

| Simulation | Cell Size ($dx_0$) | Time Step ($dt_0$) |
|---|---|---|
| **Explicit** | 0.5, 1 | 1 |
| **Direct Implicit** | 0.5, 1, 2, 3 | 1, 2, 5, 10 |
| **Energy Conserving** | 0.5, 1, 2, 3 | 1, 2, 5, 10 |

3.2 Simulation Results

Both the direct implicit (DI) and energy conserving (EC) algorithms can offer improvements in simulation time. We show a comparison in computational hours between DI-PIC simulations and MC-PIC simulations in Fig. 5. Each of these simulations is performed on the ANTYA HPC facility at the Institute for Plasma Research in India with 1,440 cores. As we can see, when comparing explicit MC-PIC, DI-PIC, and EC-PIC with the same time step and cell size, the computational cost is similar. Improvements become apparent when considering larger cell sizes for EC-PIC and DI-PIC, where adopting cell sizes $dx/dx_0 > 1$ can allow for order of magnitude reductions in compute time. This would not be possible with MC-PIC due to the emergence of the finite-grid instability under these conditions. DI-PIC also allows for adopting a larger time step $dt/dt_0 > 1$, allowing for more than an order of magnitude reduction in runtime. Therefore, using larger time step ($dt$) and cell size ($dx$) in DI-PIC and EC-PIC allows for significantly faster simulations, which is of benefit to future industrial applications. However, the computational efficiency does not increase proportionally with the increase in time step and cell size. This is mainly due to the equal need for outputting diagnostics, which impose some computational time nearly independent of time step and cell size.

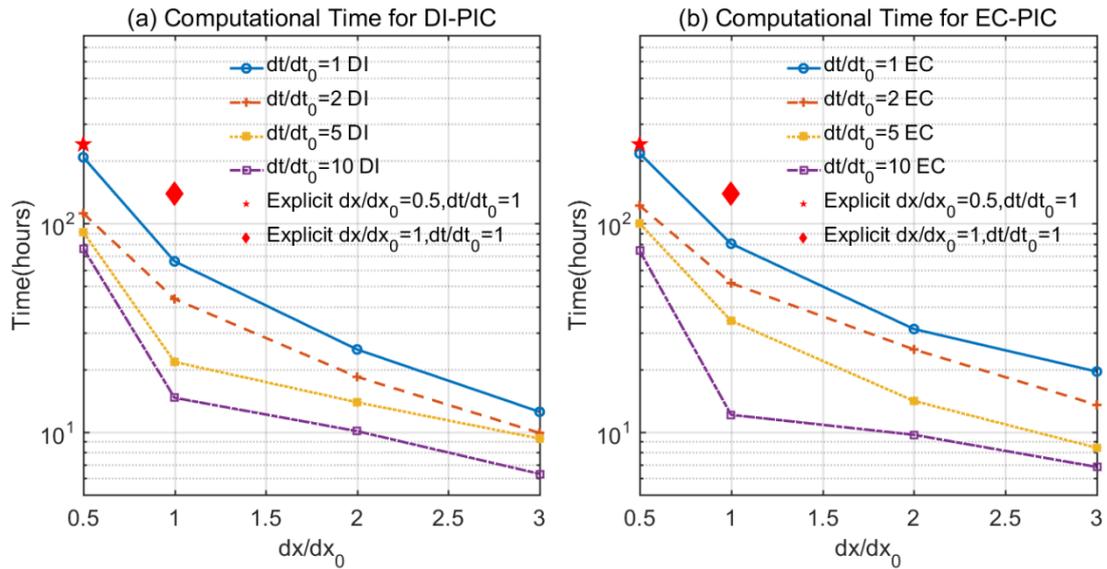

**Figure 5**: Comparison of computational time between: (a) MC-PIC (denoted by the red symbols) and DI-PIC (denoted by the lines) and (b) MC-PIC and EC-PIC. Both DI-PIC and EC-PIC can offer significant computation speedup.

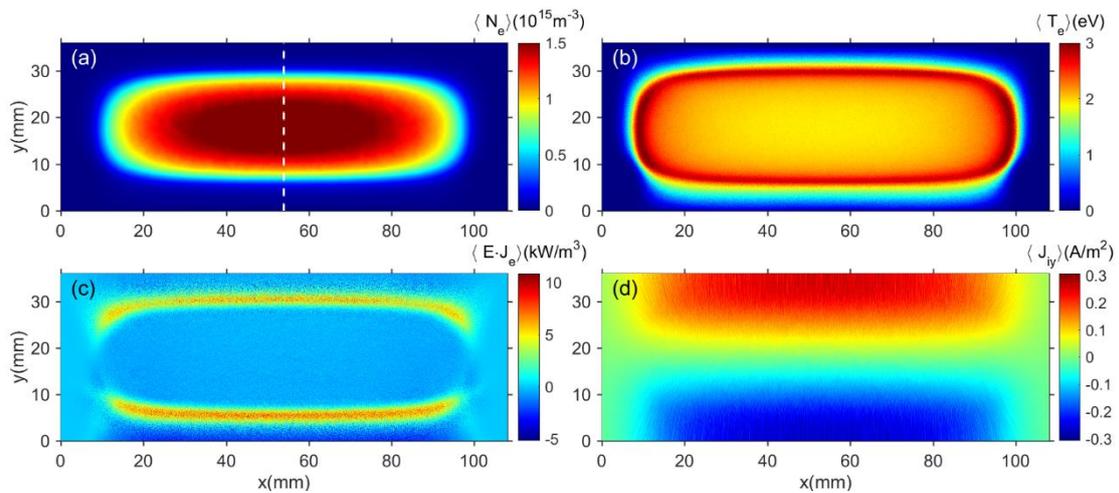

**Figure 6**: 2D profiles from the explicit MC-PIC simulation for (a) averaged electron density $\langle N_e \rangle (10^{15}m^{-3})$, (b) averaged electron temperature $\langle T_e \rangle (eV)$, (c) averaged energy

deposition to electrons $\langle E \cdot J_e \rangle (kW/m^3)$, (d) averaged ion current $\langle J_{iy} \rangle (A/m^2)$. The averages are taken over 30 rf periods. In the figures below, all the 1D cuts are taken along the white dashed line shown in (a) for the corresponding simulations.

To demonstrate the accuracy of EC-PIC and DI-PIC, we compare important physical parameters with the reference simulations performed with MC-PIC. Figure 6 (a), (b), (c) and (d) show profiles of time averaged electron density, electron temperature, electron power absorption and ion current respectively of the 2D reference simulations. All data is averaged over more than 30 rf cycles. The electron density profile ($\langle N_e \rangle$) is peaked at the center ($\sim 1.6 \times 10^{15} \; m^{-3}$) although the electron temperature ($\langle T_e \rangle$) is low at the center of discharge chamber ($\sim 2 \; eV$) and increases towards the boundaries. Temperature peaks in the vicinity of the sheath region ($\sim 3.3 \; eV$) and further decreases towards the wall. The electron power absorption ($\langle E.J_e \rangle$) is higher near to the powered electrode (bottom edge) compared to grounded electrode (top and side edges) and is asymmetric in nature (see Fig. 6 (c)). Figure 6 (d) clearly shows that the ion current is increasing towards the bottom electrode (i.e., powered) and the top electrode (i.e., grounded). In the following figures we compare the algorithms via 1D slices, with all data taking along the vertical centerline (from powered to grounded electrode) as shown in Fig. 6 (a).

Figure 7 shows the comparison of time averaged ion current for different cases. In Fig. 7 (a), (b) and (c), for DI-PIC, the values of $dx$ is varied from $dx_0$, $2dx_0$ and $3dx_0$, respectively, for specific values of $dt$ (i.e., $dt_0$, $2dt_0$, $5dt_0$ and $10dt_0$). It is important to note that the instances involving $0.5dx_0$ for both DI-PIC and EC-PIC are excluded from display because, although they offer greater accuracy, they are not computationally more efficient. Figure 7 (d), (e) and (f) further show the time averaged ion current for EC-PIC cases for the same values of $\Delta x$ and $\Delta t$ mentioned above. We found excellent agreement comparing the current at the cathode in DI-PIC and EC-PIC with the reference MC-PIC case, with an average deviation of about 7.7% (all simulation cases, including DI-PIC and EC-PIC). This indicates that our algorithms could correctly capture the profiles of ion flux, highly relevant to the performance in CCPs used for silicon etching. This could greatly enhance and impact the modelling of large volume plasma devices in the future. Figure 8 presents the comparison of time averaged energy deposition to the electrons $\langle \boldsymbol{E} \cdot \boldsymbol{J}_e \rangle_t$. The integrated energy deposition in the plot for the explicit case is $\langle \boldsymbol{E} \cdot \boldsymbol{J}_e \rangle_t = 1.92 \times 10^4 \ W/m^2$ (integrated over the 1D lineout). The average deviation of DI-PIC and EC-PIC is 21.1 % and 12.9 % respectively. Almost all EC-PIC cases (i.e. (d), (e) and (f)) provide reasonable agreement to the MC-PIC (shown in blue), while DI-PIC cases (i.e. (a), (b) and (c)) with large time steps show poorer performance. This is because at larger time steps the DI-PIC algorithm poorly resolves the time scale of the sheath motion.

Figure 9 gives a comparison of time averaged steady state electron density for different cases, i.e., for DI-PIC ((a), (b) and (c)) and for EC-PIC ((d), (e), and (f)). As we can see, almost all the cases provide good agreement with the reference MC-PIC case, with an average deviation of electron density at the center of simulation domain of only 10.3% and 8.7% for DI-PIC and EC-PIC, respectively, strongly indicating the reliability of these methods. Figure 10 further shows the comparison of time averaged electron temperature. Figure 10 (a), (b) and (c) are for the DI-PIC cases and (d), (e) and (f) are for EC-PIC cases. The temperature at center of discharge for the explicit case is nearly $2 \ eV$. In this case the temperature appears to deviate more markedly, the average deviation at the center of simulation domain is 32.1% . However, excellent agreement is found for cases with $(dt, dx) = (2dt_0, dx_0)$ i.e. in Fig. 10 (a) for DI-PIC and (d) for EC-PIC. We speculate that the lower agreement for cases with higher $(dt, dx)$ is caused by artificial numerical heating in the sheath boundary[108-110]. On the other hand, however, previous work has shown that the deviation of temperature profile does not significantly affect the accuracy of the model when compared to experiments [111, 112], this is because the electron temperature profile depends sensitively on the details of cross-section description and dynamics of fast electrons, which may not be important for the slower ion dynamics. This explains why we have an excellent match of ion current profile despite a relatively poorer agreement in electron temperature.

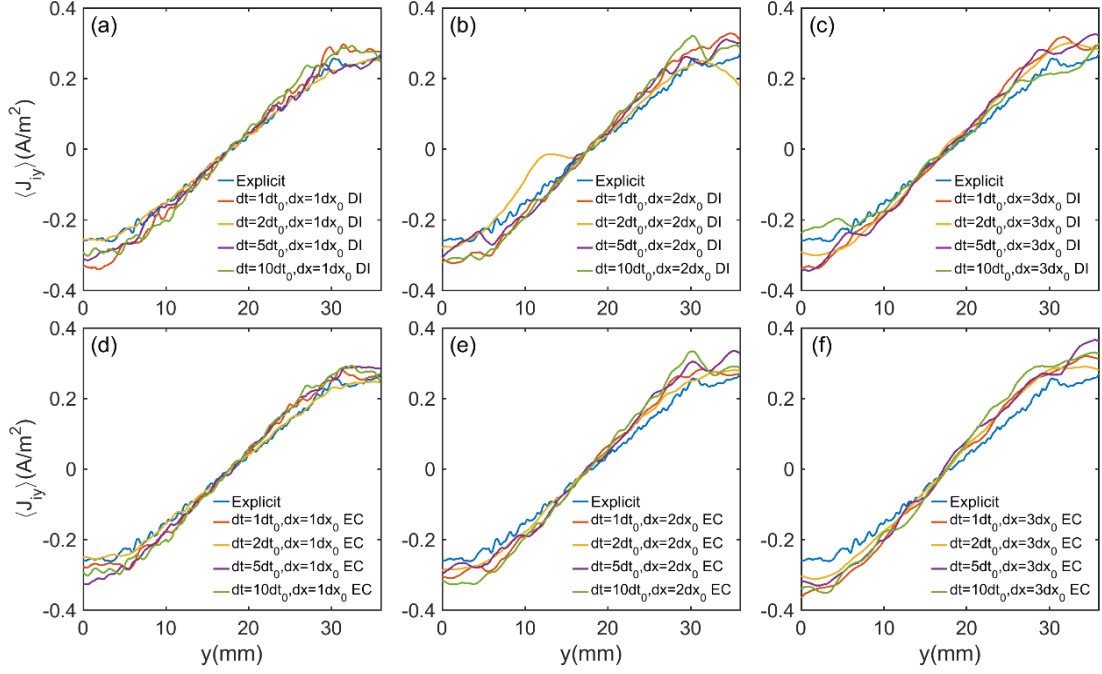

**Figure 7**: Comparison of time averaged ion current in $y$ direction $\langle J_{iy} \rangle (A/m^2)$ between explicit MC-PIC case and (a) direct implicit (DI) cases with $dx = 1dx_0$, (b) direct implicit (DI) cases with $dx = 2dx_0$, (c) direct implicit (DI) cases with $dx = 3dx_0$, (d) energy conserving (EC) cases with $dx = 1dx_0$, (e) energy conserving (EC) cases with $dx = 2dx_0$, (f) energy conserving (EC) cases with $dx = 3dx_0$. The time averages are taken at the vertical centerline over more than 30 rf cycles to reduce the noise error caused by numerical computation.

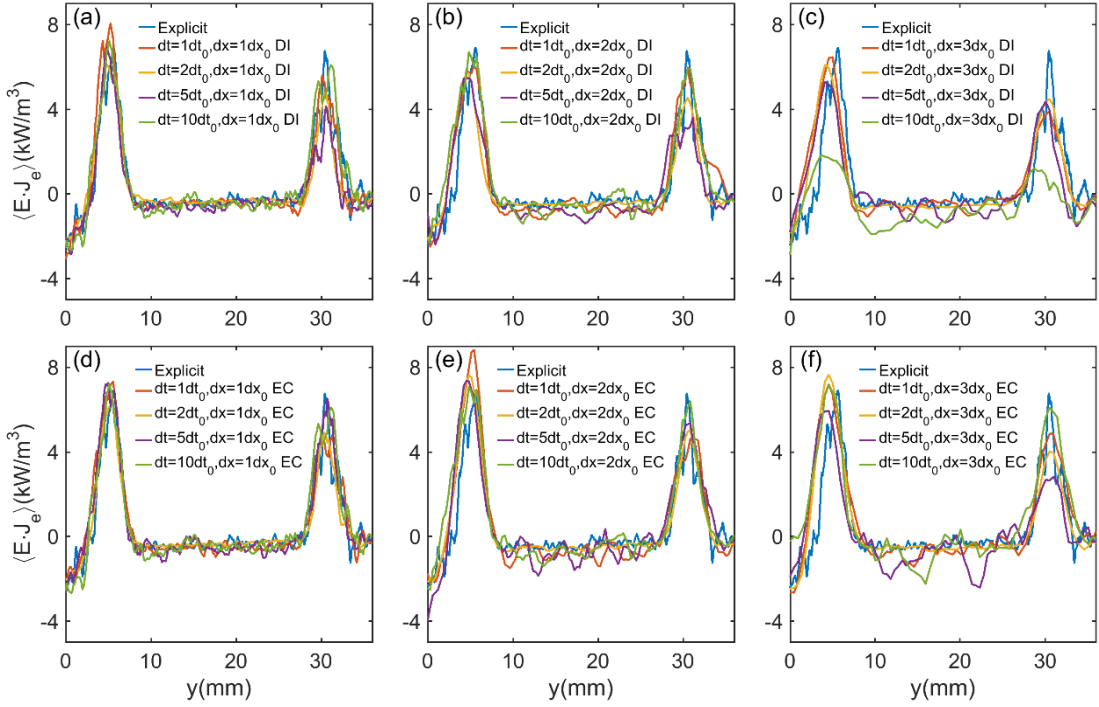

**Figure 8**: Comparison of averaged energy deposition to electrons $\langle \boldsymbol{E} \cdot \boldsymbol{J}_e \rangle (kW/m^3)$ between explicit MC-PIC case and (a) direct implicit (DI) cases with $dx = 1dx_0$, (b) direct implicit (DI) cases with $dx = 2dx_0$, (c) direct implicit (DI) cases with $dx = 3dx_0$, (d) energy conserving (EC) cases with $dx = 1dx_0$, (e) energy conserving (EC) cases with $dx = 2dx_0$, (f) energy conserving (EC) cases with $dx = 3dx_0$. The time averages are taken at the vertical centerline over more than 30 rf cycles to reduce the noise error caused by numerical computation.

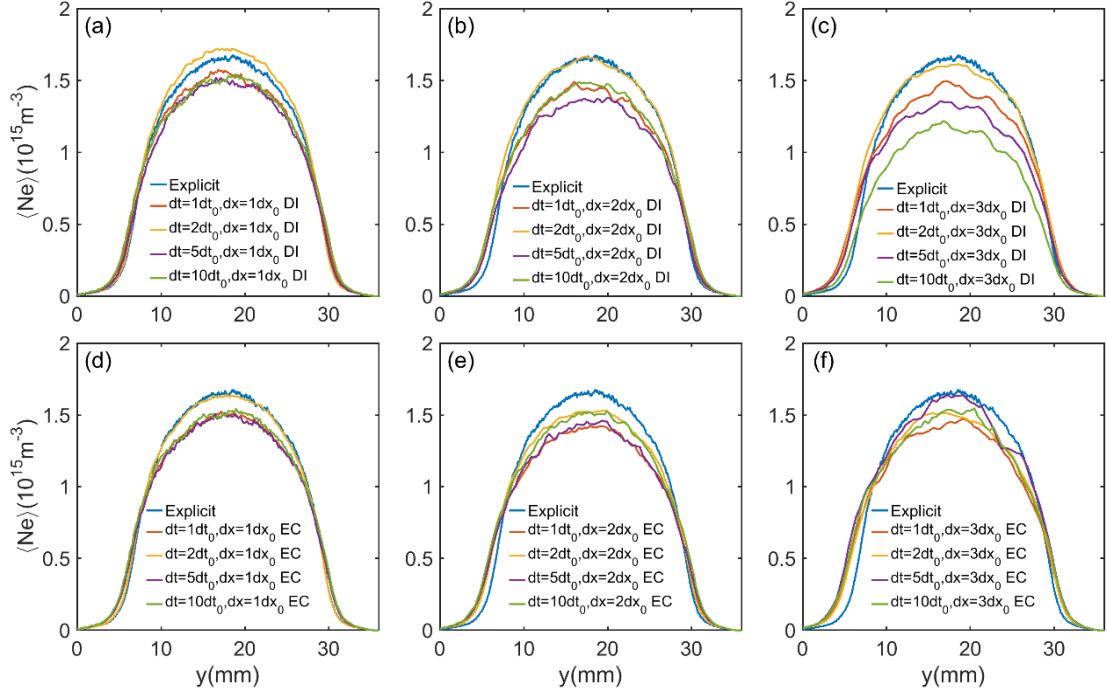

**Figure 9**: Comparison of time averaged electron density between the explicit MC-PIC case and (a) direct implicit cases with $dx = 1dx_0$, (b) direct implicit cases with $dx = 2dx_0$, (c) direct implicit cases with $dx = 3dx_0$, (d) energy conserving cases with $dx = 1dx_0$, (e) energy conserving cases with $dx = 2dx_0$, (f) energy conserving cases with $dx = 3dx_0$. The time averages are taken at the vertical centerline over more than 30 rf cycle to reduce the noise error caused by numerical computation.

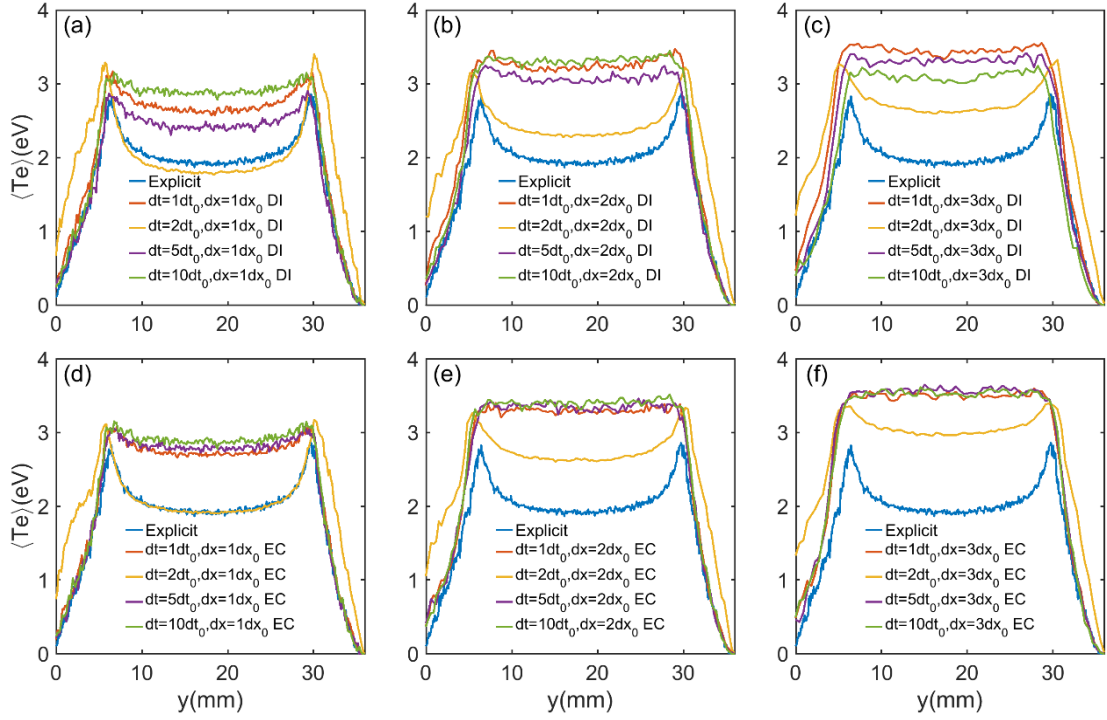

**Figure 10**: Comparison of time averaged electron temperature between the explicit MC-PIC case and (a) direct implicit cases with $dx = 1dx_0$, (b) direct implicit cases with $dx = 2dx_0$, (c) direct implicit cases with $dx = 3dx_0$, (d) energy conserving cases with $dx = 1dx_0$, (e) energy conserving cases with $dx = 2dx_0$, (f) energy conserving cases with $dx = 3dx_0$. The time averages are taken at the vertical centerline over more than 30 rf cycles to reduce the noise error caused by numerical computation.

The time evolution of different physical quantities is also compared via probe diagnostics. For each simulation, the probe is placed at the center of the simulation domain. Figures 11-12 show the probe diagnostics for ion density ($N_i$) and electron temperature ($T_e$), respectively. As we can see, the ion density (see Fig. 11) matches well with the explicit MC-PIC case (shown in blue), while time evolution of electron temperature (see Fig. 12) deviates, showing how numerical cooling or heating can play a role in determining the electron temperature in the simulations. For similar reasons, the deviation of temperature evolution is acceptable and may not affect plasma parameters relevant to real plasma devices [111, 112].

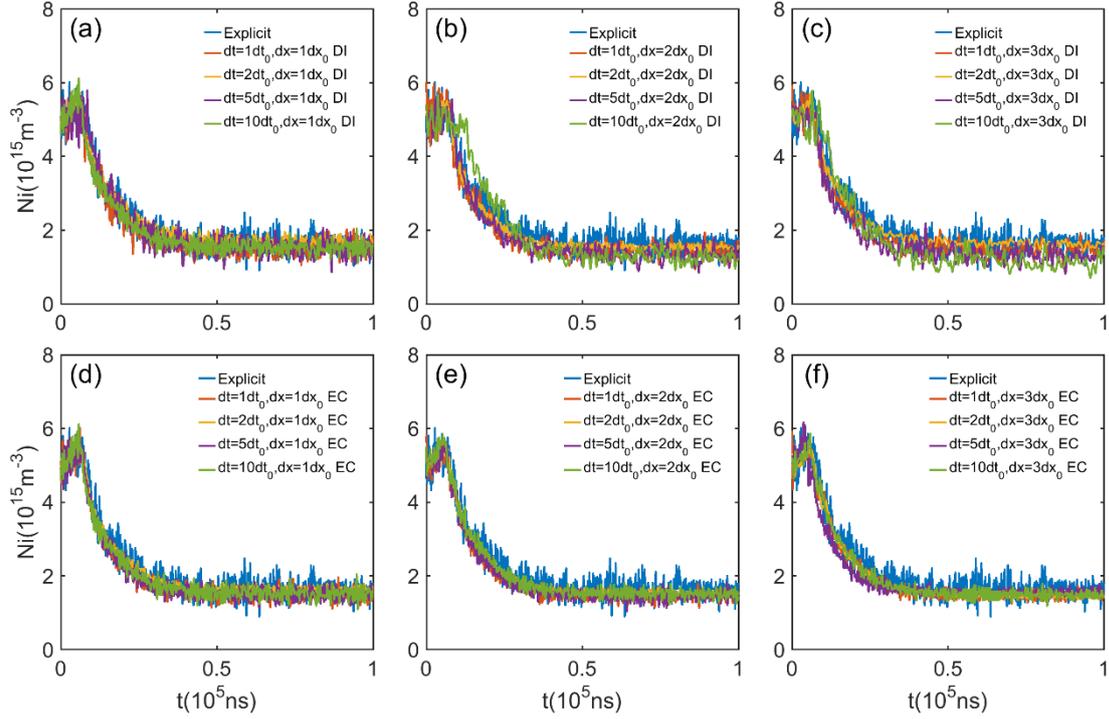

**Figure 11**: Probe diagnostics of time evolution of ion density between the explicit MC-PIC case and (a) direct implicit (DI) cases with $dx = 1dx_0$, (b) direct implicit (DI) cases with $dx = 2dx_0$, (c) direct implicit (DI) cases with $dx = 3dx_0$, (d) energy conserving (EC) cases with $dx = 1dx_0$, (e) energy conserving (EC) cases with $dx = 2dx_0$, (f) energy conserving (EC) cases with $dx = 3dx_0$. The probes are placed at the center of simulation domain.

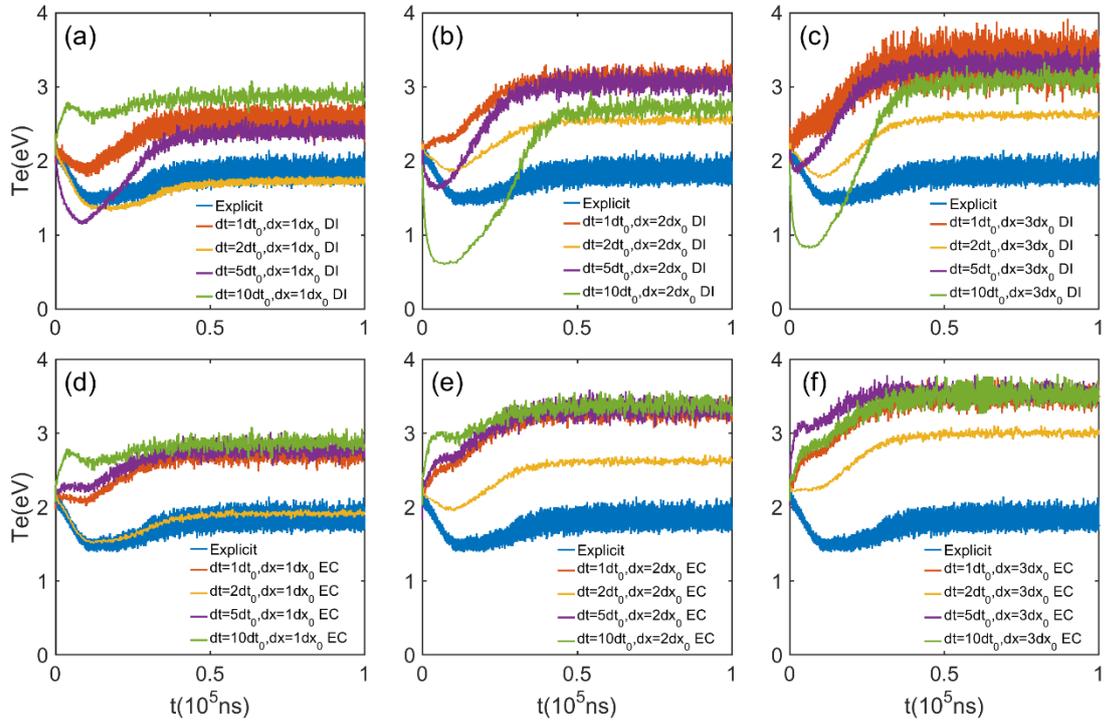

**Figure 12**: Probe diagnostics of time evolution of electron temperature between the explicit MC-PIC case and (a) direct implicit (DI) cases with $dx = 1dx_0$, (b) direct implicit (DI) cases with $dx = 2dx_0$, (c) direct implicit (DI) cases with $dx = 3dx_0$, (d) energy conserving (EC) cases with $dx = 1dx_0$, (e) energy conserving (EC) cases with $dx = 2dx_0$, (f) energy conserving (EC) cases with $dx = 3dx_0$. The probes are placed at the center of simulation domain.

The error analysis mentioned above is further summarized in Fig. 13, where not only the average error for each physical quantity but also the error in each simulation case is shown. Close to the optimal path for DI-PIC (corresponding to $dx/dx_0 = 1$, $dt/dt_0 = 2$), the error is significantly smaller. It is also shown that the error in general increases with the increase in time step and cell size, however the increment is not significant. Therefore, our codes not only maintain stability under large time step and cell size, but also have acceptable numerical accuracy, which strongly benefits the future industrial usage. Figure 14 shows the Electron Velocity Distribution Function (EVDF) for several representative simulations. We can see that all of them are non-Maxwellian, and the EVDFs for both DI-PIC and EC-PIC do not deviate significantly from the explicit case. Despite some small deviations in the distribution of energetic electrons shown in Fig. 14 (b), the similarity in ion flux shown in Fig. 7 indicates that this has a small effect on the ions impacting the electrode. Therefore, although there is a small compromise of numerical accuracy within our code when modelling kinetic effects, this compromise is generally deemed acceptable. A more detailed analysis for such a numerical distortion will be investigated in future publications.

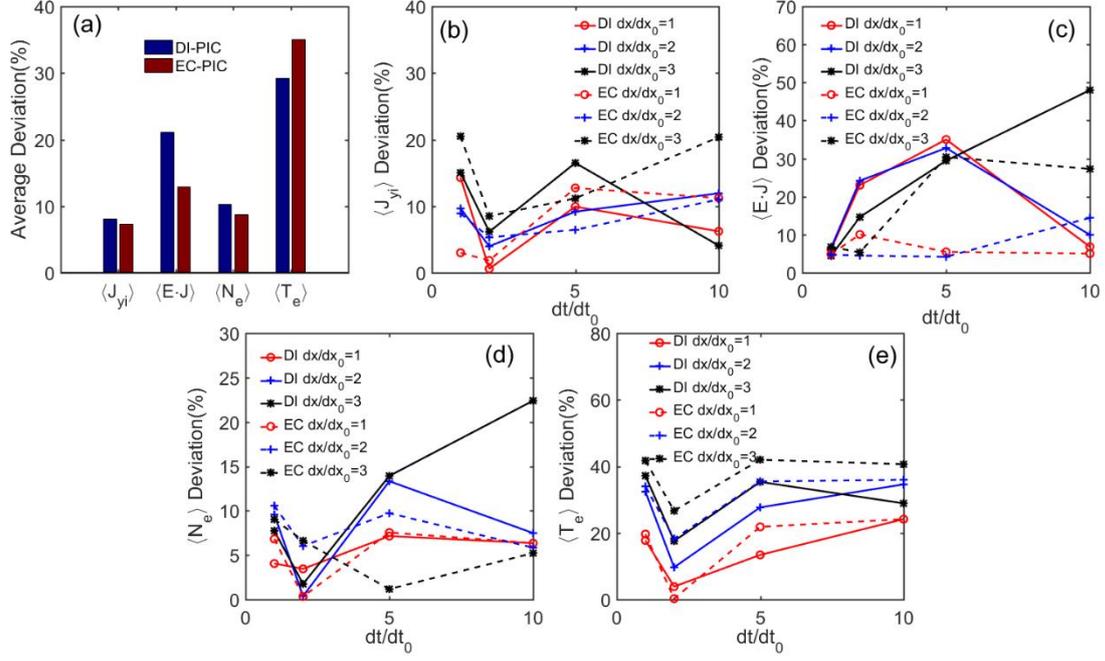

**Figure 13**: Summary of error analysis. (a) shows the average deviation of the four physical quantities in all the simulations. (b)-(e) gives the details of the deviation as a function of time step and cell size, where solid lines are for DI-PIC and dotted lines are for EC-PIC.

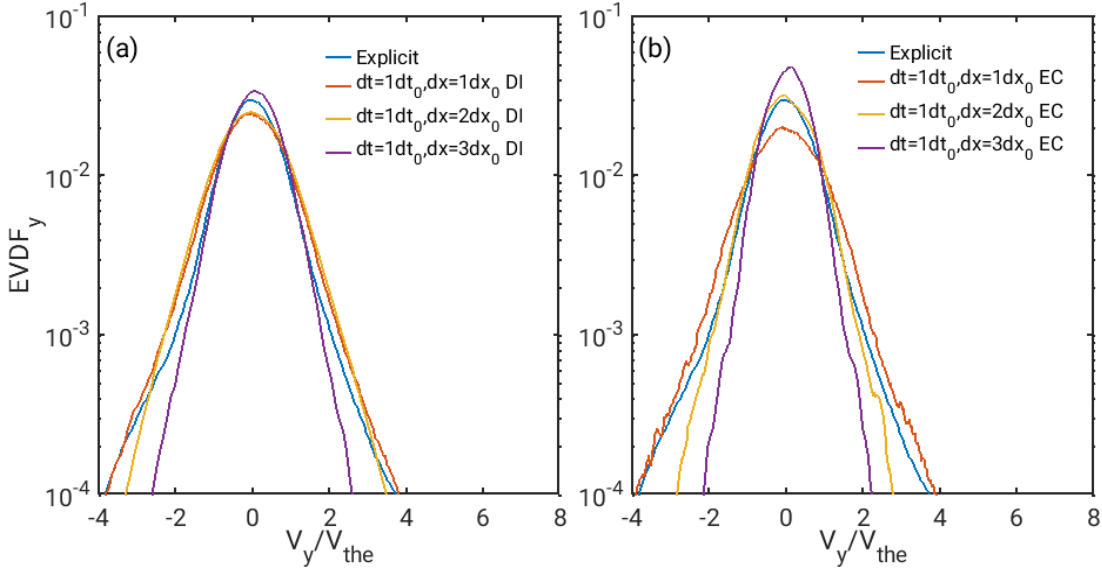

**Figure 14**: Electron Velocity Distribution Function (EVDF) comparison between the explicit MC-PIC case and (a) direct implicit (DI) cases with $dt = 1dt_0$, (b) energy conserving (EC) cases with $dt = 3dt_0$. All the EVDFs are taken from the same region at the center of the simulation domain. $V_{the} = \sqrt{2T_{e0}/m_e}$ is the initial thermal speed of electrons.

### 3.3 Numerical heating or cooling of direct implicit (DI) algorithm

As was established in Sec. 2, the numerical heating of DI has a linear dependence on the ratio of time step and cell size i.e., $dt/dx$. Such a dependence has only been tested in a double periodic system, not in a practical configuration with electrodes [85]. Here, in Fig. 15, we show that such a linear dependence persists when time the step is small (shown in Fig. 15 (a)), but fails to hold when $dt > 5dt_0$ (shown in Fig. 15 (b)). However, as shown in Fig. 16, the linear dependence persists when the rf voltage amplitude is increased to $V_{rf} = 1000\ V$, although analyzing such complex behavior of numerical heating is beyond the scope of this paper. Given a certain time step and cell size, one could estimate the numerical heating using a simple linear curve shown as shown in Fig. 15 (a) and Fig. 16, with a caveat that this will no longer be valid for very large time steps. On the other hand, as is shown in Fig. 5, a significant improvement in computational efficiency compared to traditional explicit MC-PIC is already achieved even when using a relatively small time step increase, say $dt = 2dt_0$.

This implies that, despite the relative inaccuracy in estimating electron temperature compared with more accurate, but expensive, algorithms (for instance, fully implicit algorithms [92, 93]), the numerical heating of the direct implicit algorithm is acceptable and could be directly implemented into simulations for modeling practical plasma devices, or be used to gain an insight into the physical trends before conducting full resolution simulations.

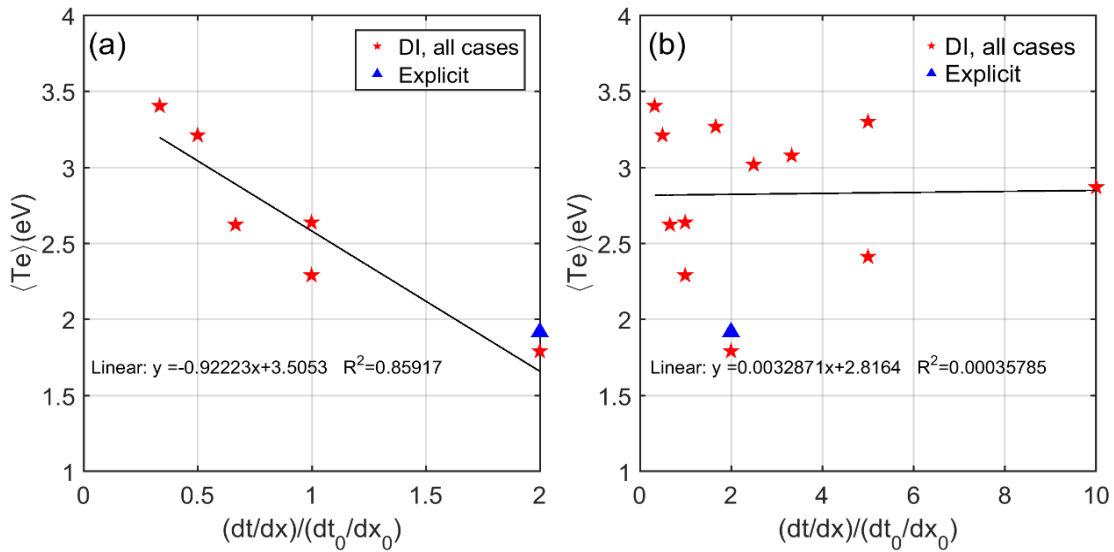

**Figure 15**: Scatter plot showing (a) averaged electron temperature as a function of $dt/dx$ for MC-PIC case and DI-PIC cases only with small time steps ($dt = 1dt_0$ and $dt = 2dt_0$) and (b) averaged electron temperature as a function of $dt/dx$ for all the DI-PIC cases listed in Table 3. A good linear fit for temperature occurs for relatively small time steps.

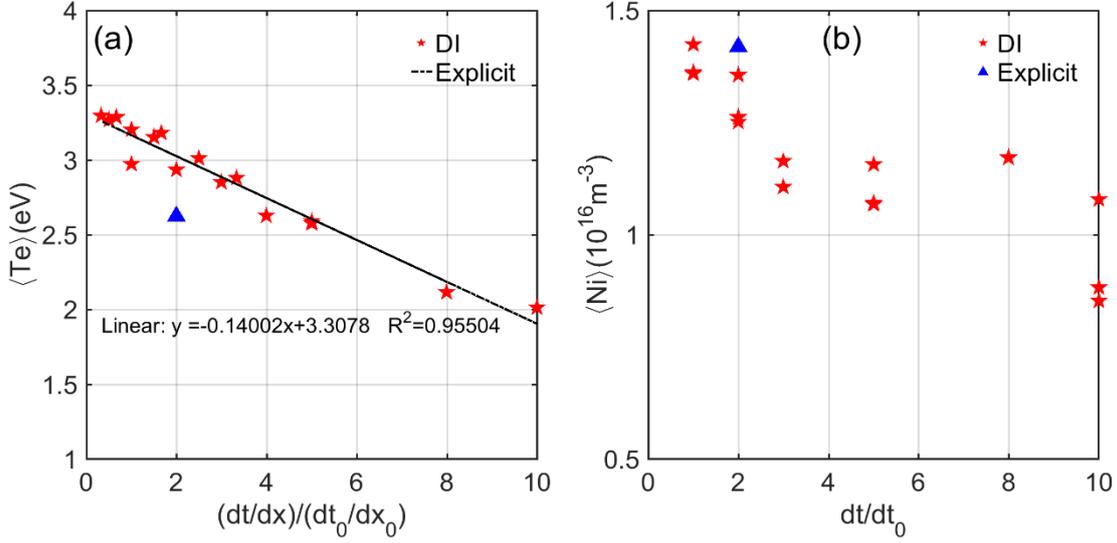

**Figure 16**: Scatter plot showing (a) averaged electron temperature as a function of $dt/dx$ and (b) averaged ion density as a function of $dt$ for several other DI-PIC cases with $V_{rf} = 1000V$ (other parameters keeping the same). A good linear fit for temperature occurs for all the cases.

### 3.4 Benchmarking against Low-Temperature Plasma Particle-in-Cell (LTP-PIC) code

To gain more confidence in our results, we benchmarked our simulations for different cases against a 2D version of the LTP-PIC code. The Low-Temperature Plasma Particle-in-Cell (LTP-PIC) code is a general-purpose high-performance 3D-3V kinetic plasma simulation software designed and developed at the Princeton Plasma Physics Laboratory (PPPL)[113-115]. The energy conserving version of LTP-PIC (implementing the same algorithm as discussed in Section 2.2) can also simulate plasma devices with cell size much greater than the electron Debye length. At the same time, LTP-PIC uses the similar collision algorithms as EDIPIC-2D. Therefore, a direct comparison between simulation results from the two codes is possible.

Here, we consider four representative cases for the benchmark exercise, which are the explicit cases ($dt = 0.5dt_0$, $dx = 0.5dx_0$) (both EDIPIC-2D and LTP-PIC) and then using EC-PIC (EDIPIC-2D and LTP-PIC) and DI-PIC (EDIPIC-2D only) with $dt = 5dt_0$, $dx = 1, 2, 3dx_0$. Figure 17-20 show the profile benchmark results for time average ion current ($\langle J_{iy} \rangle$), time averaged energy deposition ($\langle E \cdot J_e \rangle$), time average electron density ($\langle N_e \rangle$) and time average electron temperature ($\langle T_e \rangle$), respectively. For all four physical quantities, the deviations of explicit MC-PIC from LTP-PIC (shown by subfigures (a) in Fig. 17-20) are within 8%. For ion current at the cathode, DI-PIC and EC-PIC deviate from LTP-PIC by about 8.9%. For the sum of 1D energy deposition, DI-PIC and EC-PIC deviate from LTP-PIC by about 18.1%. For electron density at the simulation domain center, the average deviation of DI-PIC and EC-PIC from LTP-PIC is about 6.6%. For electron temperature at the domain center, the average deviation of DI-PIC and EC-PIC from LTP-PIC is about 6.9%. Therefore, all the cases exhibit a satisfactory level of agreement and deviations are believed to arise from subtle disparities in the Monte-Carlo collision algorithms employed by each code.

Figure 21 further benchmarked the probe diagnostics for ion density ($N_i$), and excellent agreement is found. In conclusion, both DI-PIC and EC-PIC code are successfully benchmarked and validated, which could be directly used in future large volume plasma device modelling.

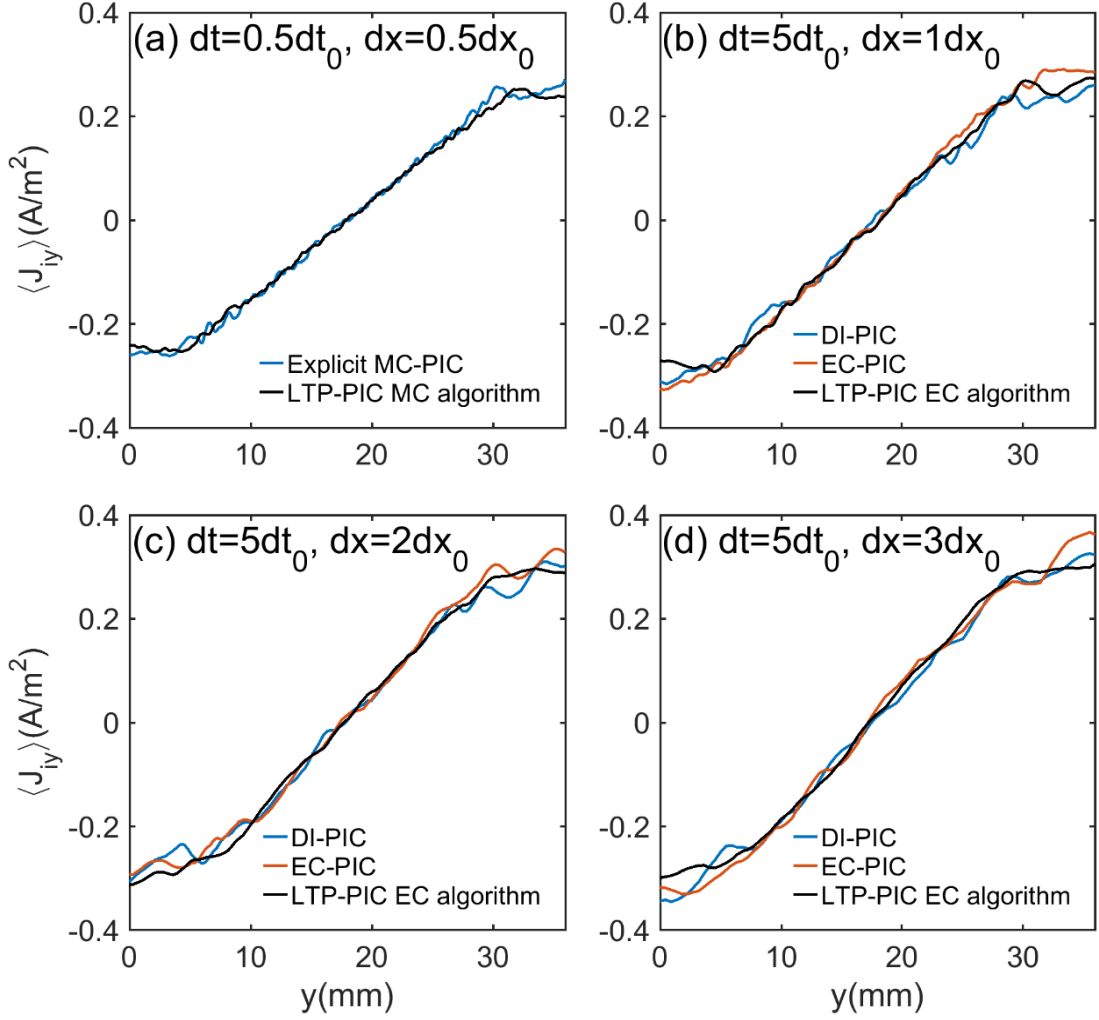

**Figure 17**: Average ion current benchmark of EDIPIC-2D with LTP-PIC code for (a) explicit MC-PIC code, (b) DI-PIC code and EC-PIC code with $dt = 5dt_0, dx = 1dx_0$, (c) DI-PIC code and EC-PIC code with $dt = 5dt_0, dx = 2dx_0$, (d) DI-PIC code and EC-PIC code with $dt = 5dt_0, dx = 3dx_0$. The time averages are taken over more than 30 rf cycles to reduce the noise error caused by numerical computation. Note that the LTP-PIC benchmark for (a) is using the MC-PIC algorithm while LTP-PIC benchmark for (b)-(d) are using the EC-PIC algorithm.

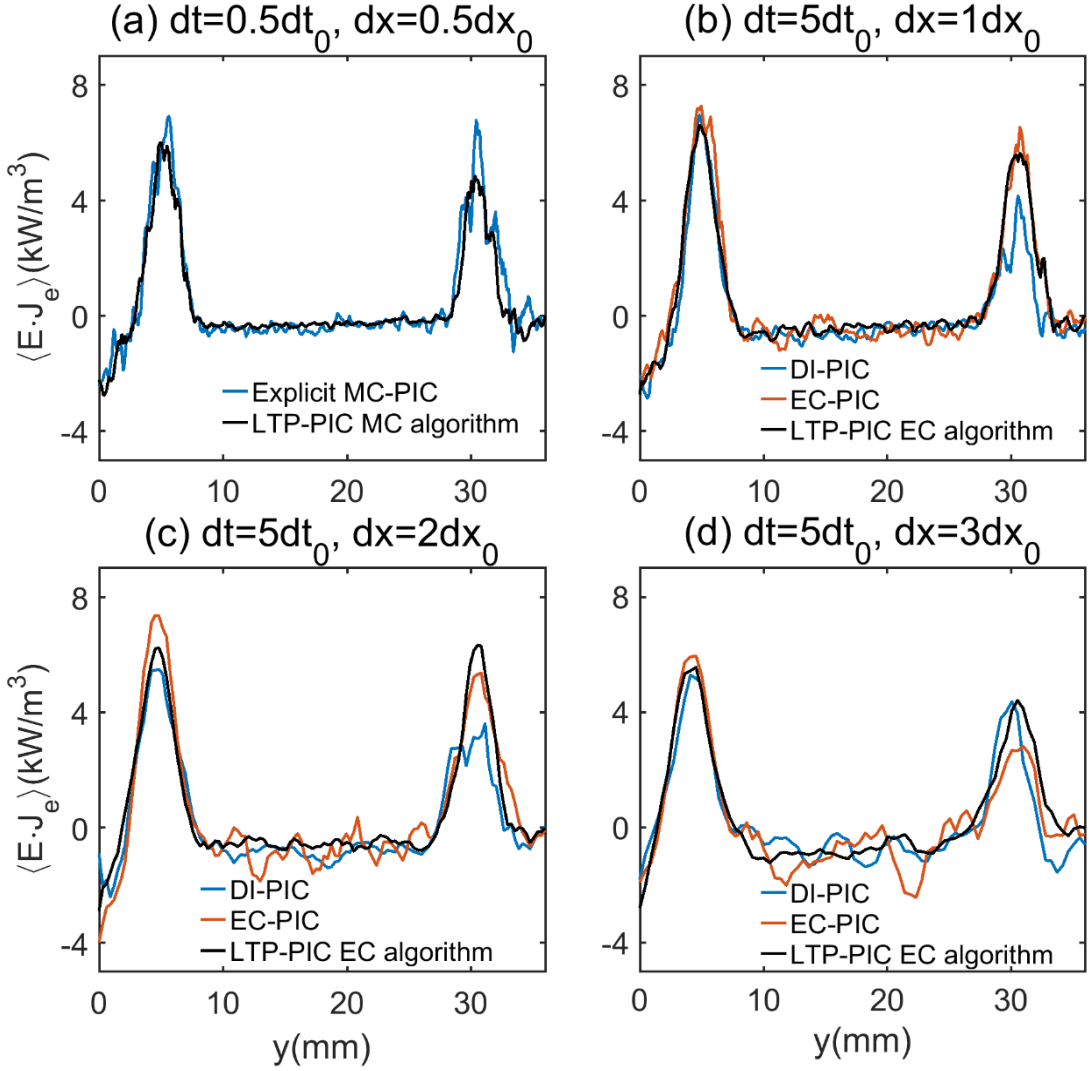

**Figure 18**: Average electron energy deposition benchmark of EDIPIC-2D with LTP-PIC code for (a) explicit MC-PIC code, (b) DI-PIC code and EC-PIC code with $dt = 5dt_0, dx = 1dx_0$, (c) DI-PIC code and EC-PIC code with $dt = 5dt_0, dx = 2dx_0$, (d) DI-PIC code and EC-PIC code with $dt = 5dt_0, dx = 3dx_0$. The time averages are taken over more than 30 rf cycles to reduce the noise error caused by numerical computation. Note that the LTP-PIC benchmark for (a) is using the MC-PIC algorithm while LTP-PIC benchmark for (b)-(d) are using the EC-PIC algorithm.

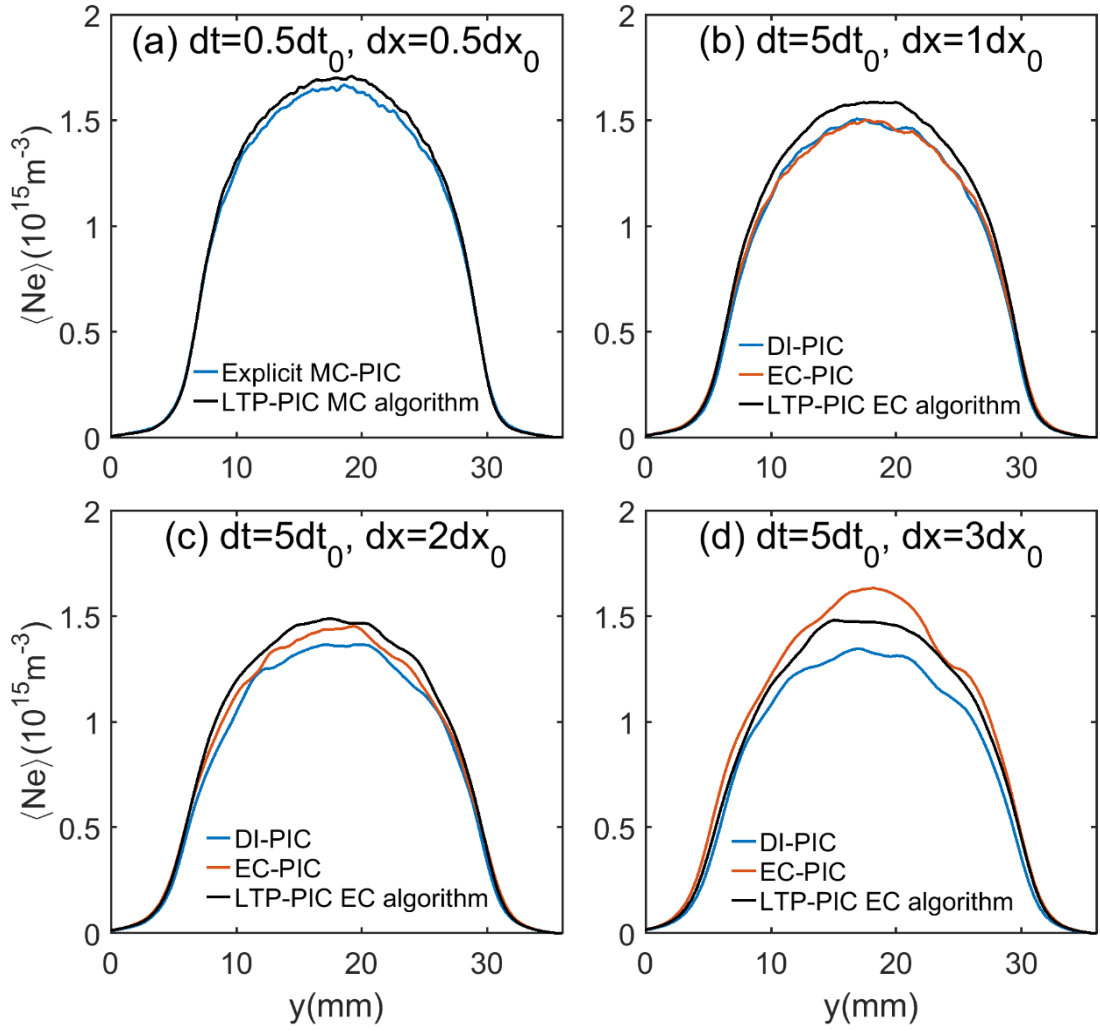

**Figure 19**: Average electron density benchmark of EDIPIC-2D with LTP-PIC code for (a) explicit MC-PIC codes, (b) DI-PIC code and EC-PIC code with $dt = 5dt_0, dx = 1dx_0$, (c) DI-PIC code and EC-PIC code with $dt = 5dt_0, dx = 2dx_0$, (d) DI-PIC code and EC-PIC code with $dt = 5dt_0, dx = 3dx_0$. The time averages are taken over more than 30 rf cycles to reduce the noise error caused by numerical computation. Note that the LTP-PIC benchmark for (a) is using the MC-PIC algorithm while LTP-PIC benchmark for (b)-(d) are using the EC-PIC algorithm.

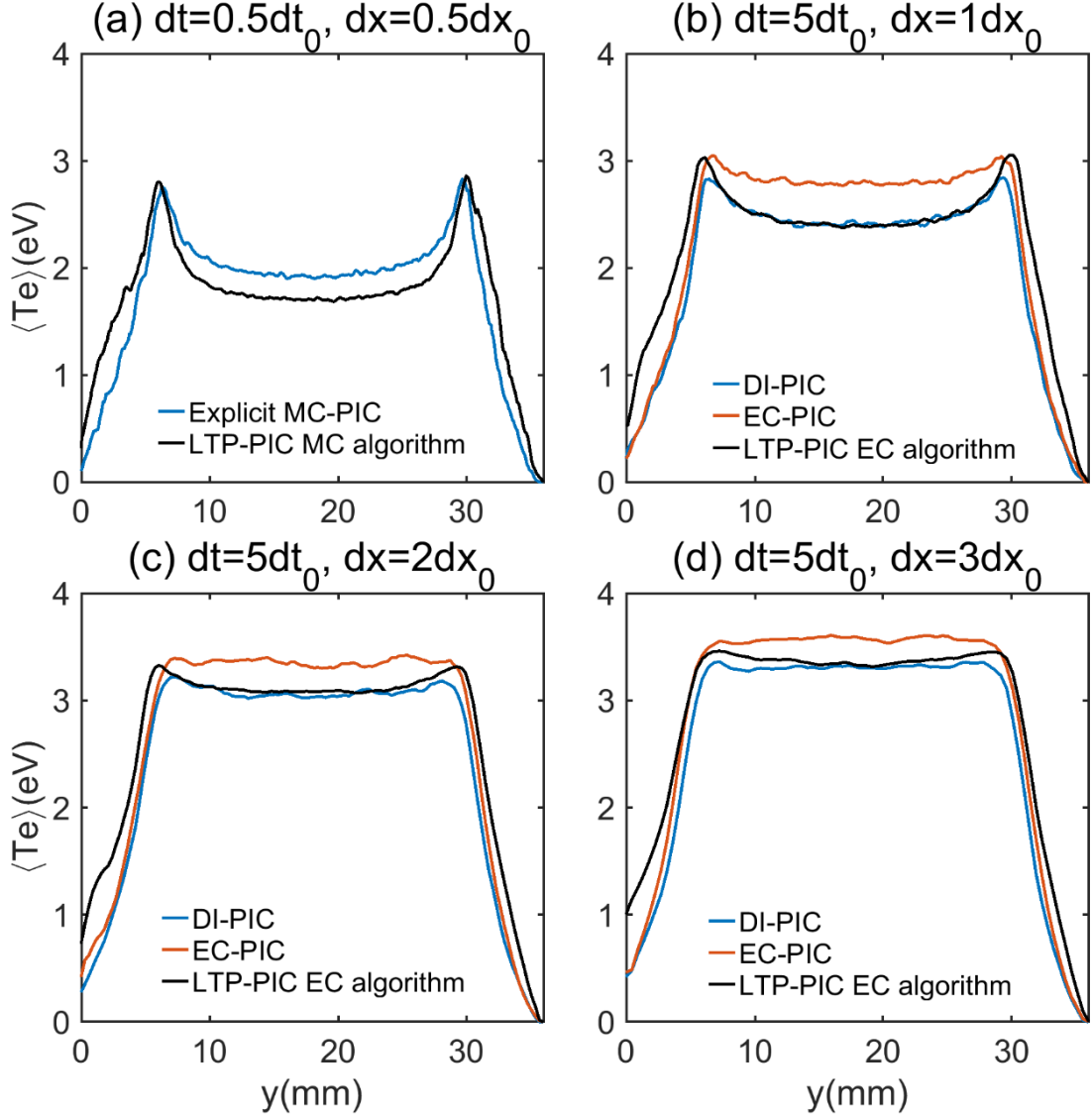

**Figure 20**: Average electron temperature benchmark of EDIPIC-2D with LTP-PIC code for (a) explicit MC-PIC codes, (b) DI-PIC code and EC-PIC code with $dt = 5dt_0, dx = 1dx_0$, (c) DI-PIC code and EC-PIC code with $dt = 5dt_0, dx = 2dx_0$, (d) DI-PIC code and EC-PIC code with $dt = 5dt_0, dx = 3dx_0$. The time averages are taken over more than 30 rf cycles to reduce the noise error caused by numerical computation. Note that the LTP-PIC benchmark for (a) is using the MC-PIC algorithm while LTP-PIC benchmark for (b)-(d) are using the EC-PIC algorithm.

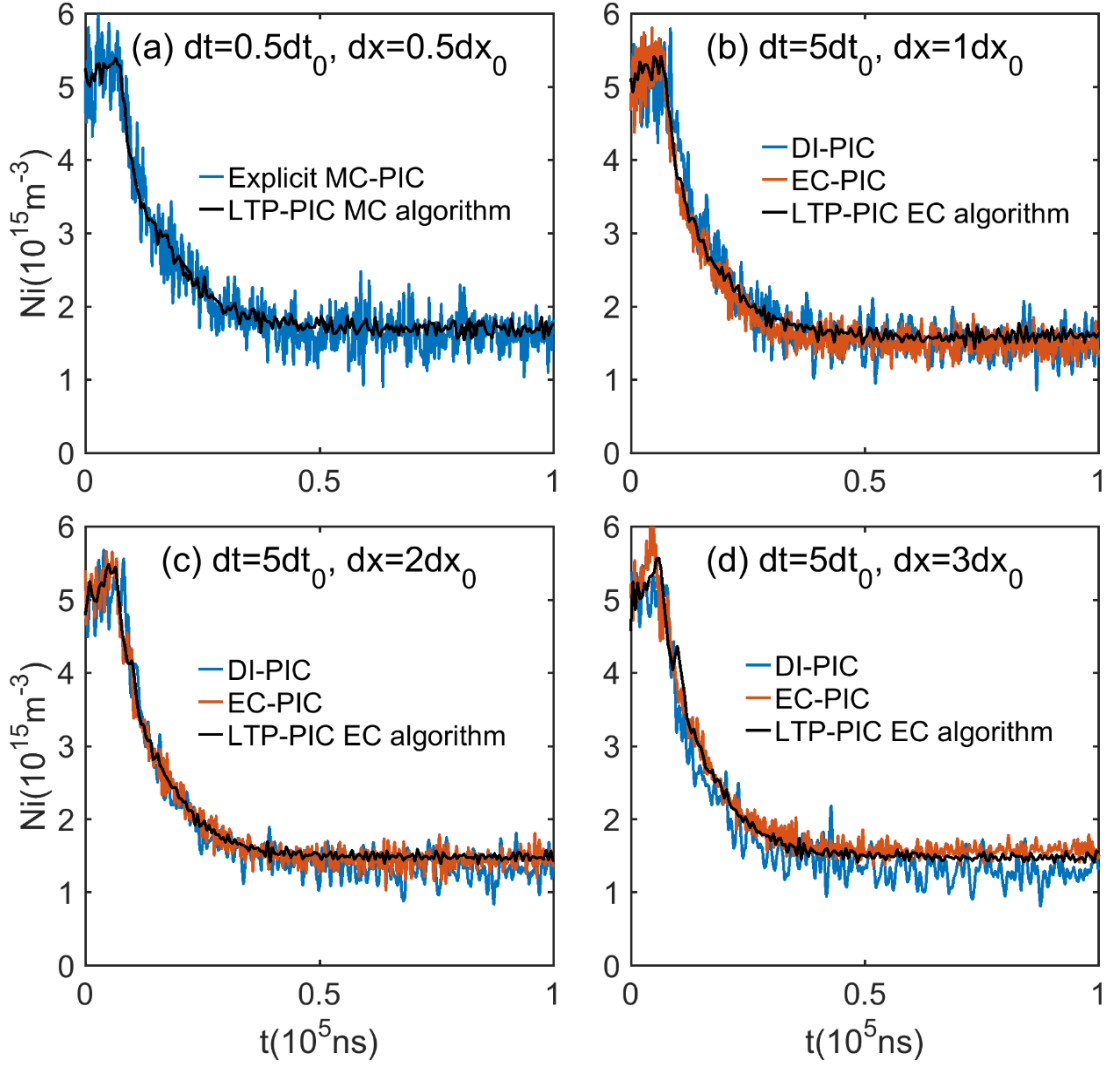

**Figure 21**: Ion density probe diagnostics benchmark between our new codes and LTP-PIC code for (a) explicit MC-PIC codes, (b) DI-PIC code and EC-PIC code with $dt = 5dt_0, dx = 1dx_0$, (c) DI-PIC code and EC-PIC code with $dt = 5dt_0, dx = 2dx_0$, (d) DI-PIC code and EC-PIC code with $dt = 5dt_0, dx = 3dx_0$. The probes are placed at the center of simulation domain.

## 4. Conclusions and Discussion

In this paper, we described the Direct Implicit (DI) and explicit Energy Conserving (EC) Particle-in-Cell algorithms, which offer a pathway to reduce computational cost of 2D simulations of modern low-temperature large volume plasma devices. We provide a comprehensive explanation of the derivations of the two algorithms and implemented them into two different PIC codes i.e., EDIPIC-2D and the LTP-PIC code. These algorithms were tested against fully resolved traditional MC-PIC simulations of a CCP discharge operated with radio-frequency. Good agreement was found between the algorithms when the time step was kept below that required to resolve the electron plasma period, even when the Debye length is under-resolved. Furthermore, these simulations were accomplished in significantly less time due to the smaller number of total cells and macro-particles. When the time step was increased, agreement was still favorable, however numerical heating/cooling led to larger deviations in electron temperature. In order to address this issue, we demonstrate a way to estimate this numerical heating, which might be applicable in even in more complex CCP configurations. The outcome of this investigation has shown that the two approaches could be used to perform significantly faster simulations, with only a minimal cost in accuracy.

We noticed that the DI algorithm, despite being very useful, does not accurately conserve energy, therefore future work will consider the use of fully implicit PIC algorithms. In this paper, we also directed our focus to the CCP discharge, however both algorithms are amenable to more complex configurations, since their implementations are able to handle magnetic fields. In future endeavors, we will also explore the application and limitations of these approaches when applied to other low-temperature plasma devices that are pertinent to material processing.

### Acknowledgement

The PPPL group was funded by the US Department of Energy through PPPL Laboratory Directed Research & Development (LDRD). Sarveshwar Sharma would like to thank ANTYA HPC facility at the Institute for Plasma Research, Gandhinagar, INDIA.

**References**

1.  M. A. Lieberman and A. J. Lichtenberg, *Principles of Plasma Discharges and Materials Processing, Chap 3.* (John Wiley & Sons, 2005).

2.  Pascal Chabert and N. Braithwaite, *Physics of Radio-frequency Plasmas*. (Cambridge University Press, 2011).

3.  Wenchong Ouyang, Chengbiao Ding, Qi Liu, Shuzhan Gao, Weifeng Deng and Zhengwei Wu, AIP Advances **11**, 075121 (2021).

4.  E. Kawamura, A. J. Lichtenberg, M. A. Lieberman and A. M. Marakhtanov, Plasma Sources Sci. Technol. **25**, 035007 (2016).

5.  Y. Yang and M. J. Kushner, Plasma Sources Sci. Technol. **19**, 055011 (2010).

6.  S. Rauf, K. Bera and K. Collins, Plasma Sources Sci. Technol. **17**, 035003 (2008).

7.  A. Agarwal, S. Rauf and K. Collins, Plasma Sources Sci. Technol. **21** (2012).

8.  E. Kawamura, M. A. Lieberman and D. B. Graves, Plasma Sources Sci. Technol. **23**, 064003 (2014).

9.  Yu-Ru Zhang, Xiang Xu and Y.-N. Wang, Phys. Plasmas **17**, 033507 (2010).

10. Na Gao, Yan-Bin Xi, Jiang-Jiang Li and Y. Liu, Vacuum **192**, 110466 (2021).

11. A. M. Velasco, J. D. Muñoz and M. Mendoza, Journal of Computational Physics **376** (76) (2019).

12. H. C. Kim, F. Iza, S. S. Yang, M. Radmilovic-Radjenovi and J. K. Lee, J. Phys. D: Appl. Phys. **38**, R283 (2005).

13. J. van Dijk, G. M. W. Kroesen and A. Bogaerts, J. Phys. D: Appl. Phys. **42**, 190301 (2009).

14. Yiting Zhang, Mark J. Kushner, Saravanapriyan Sriraman, Alexei Marakhtanov, John Holland and A. Paterson, J. Vac. Sci. Technol. **33**, 031302 (2015).

15. E. Kawamura, A. J. Lichtenberg and M. A. Lieberman, Plasma Sources Sci. Technol. **25**, 035007 (2016).

16. Haomin Sun, Yan Yang, Quanming Lu, San Lu, Minping Wan and R. Wang, the Astrophy. Journal **926:97** (97) (2022).

17. S. G. Walton, D. R. Boris, S. C. Hernandez, E. H. Lock, Tz. B. Petrova, G. M. Petrov and R. F. Fernsler, ECS Journal of Solid State Science and Technology **4** (6), N5033-N5040 (2015).

18. Stephen Muhl and A. Pérez, Thin Solid Films **579**, 174-198 (2015).

19. I. Adamovich, S. D. Baalrud, A. Bogaerts, P. J. Bruggeman and M. C. e. al., J. Phys. D: Appl. Phys. **50**, 323001 (2017).

20. R. W. Hockney and J. W. Eastwood, *Computer Simulation using Particles*. (New York: Adam Hilger, 1988).

21. C. K. Birdsall and A. B. Langdon, *Plasma Physics via Computer Simulation*. (New York: McGraw-Hill, 1985).

22. G. A. Bird, *Molecular Gas Dynamics and the Direct Simulation of Gas Flows*. (Oxford: Clarendon, 1994).

23. D. Vender and R. W. Boswell, IEEE Trans. Plasma Sci. **18**, 725 (1990).

24. M. Yan and W. J. Goedheer, Plasma Sources Sci. Technol. **8**, 349 (1999).

25. H. M. Sun and J. Sun, J. Geophys. Res. **125**, e2019JA027376 (2019).

26. I. V. Sokolov, Haomin Sun, Gabor Toth, Zhenguang Huang, Valeriy Tenishev, Lulu Zhao, Jozsef Kota, Ofer Cohen and T. Gombosi, Journal of Computational Physics **476**, 111923 (2023).

27. Haomin Sun, Jian Chen, Igor D. Kaganovich, Alexander Khrabrov and D. Sydorenko, Phys. Rev. Lett. **129**, 125001 (2022).

28. Haomin Sun, Jian Chen, Igor D. Kaganovich, Alexander Khrabrov and D. Sydorenko, Phys. Rev. E **106**, 035203 (2022).

29. San Lu, V. Angelopoulos, P. L. Pritchett, Jia Nan, Kai Huang, Xin Tao, A. V. Artemyev, A. Runov, Yingdong Jia, Haomin Sun and N. Kang, Journal of Geophysical Research: Space Physics **126**, e2021JA2020 (2021).

30. San Lu, Quanming Lu, Rongsheng Wang, Xinmin Li, Xinliang Gao, Kai Huang, Haomin Sun, Yan Yang, Anton V. Artemyev, Xin An and Y. Jia, The Astrophy. Journal **943:100** (2023).

31. Yong-Xin Liu, Quan-Zhi Zhang, Wei Jiang, Lu-Jing Hou, Xiang-Zhan Jiang, Wen-Qi Lu and Y.-N. Wang, Phys. Rev. Lett. **107**, 055002 (2011).

32. E. Kawamura, M. A. Lieberman and A. J. Lichtenberg, Phys. Plasmas **13**, 053506 (2006).

33. Sarveshwar Sharma, Sanket Patil, Sudip Sengupta, Abhijit Sen, Alexander Khrabrov and a. I. Kaganovich, Phys. Plasmas **29**, 063501 (2022).

34. Sanket Patil, Sarveshwar Sharma, Sudip Sengupta, Abhijit Sen and I. Kaganovich, Phys. Rev. Lett. **4**, 013059 (2022).

35. Shu Zhang, Guang-Yu Sun, Jian Chen, Haomin Sun, An-Bang Sun and G.-J. Zhang, Appl. Phys. Lett. **121**, 014101 (2022).

36. L. Xu, L. Chen, M. Funk, A. Ranjan, M. Hummel, R. Bravenec, R. Sundararajan, D. J. Economou and V. M. Donnelly, Appl. Phys. Lett. **93**, 261502 (2008).

37. U. Buddemeier, U. Kortshagen and I. Pukropski, Appl. Phys. Lett. **67** (191), 191 (1995).

38. V. A. Godyak, Sov. J. Plasma Phys. **2**, 78 (1976).

39. O. A. Popov and V. A. Godyak, J. Appl. Phys. **57** (53) (1985).

40. M. A. Lieberman, IEEE Trans. Plasma Sci. **16** (638), 638 (1988).

41. I. D. Kaganovich, Phys. Rev. Lett. **89**, 265006 (2002).

42. I. D. Kaganovich, O. V. Polomarov and C. E. Theodosiou, IEEE Trans. Plasma Sci. **34**, 696 (2006).

43. S. Sharma and M. M. Turner, Phys. Plasmas **20** (7), 073507 (2013).

44. S. Sharma and M. M. Turner, Plasma Sources Sci. Technol. **22**, 035014 (2013).

45. S. Sharma, S. K. Mishra and P. K. Kaw, Phys. Plasmas **21**, 073511 (2014).

46. Guang-Yu Sun, Han-Wei Li, An-Bang Sun, Yuan Li, Bai-Peng Song, Hai-Bao Mu, Xiao-Ran Li and G. J. Zhang, Plasma Processes Polymers **16** (93), 1900093 (2019).

47. Guangyu Sun, Shu Zhang, Anbang Sun and G. ZHANG, Plasma Sci. Technol. **24**, 095401 (2022).

48.    H. H. Goto, H. D. Lowe and T. Ohmi, IEEE Trans. Semicond. Manuf. **6** (58), 58 (1993).

49.    J. Robiche, P. C. Boyle, M. M. Turner and A. R. Ellingboe, J. Phys. D: Appl. Phys. **38**, 1810 (2003).

50.    H. C. Kim, J. K. Lee and J. W. Shon, Phys. Plasmas **10**, 4545 (2003).

51.    M. M. Turner and P. Chabert, Phys. Rev. Lett. **96**, 205001 (2006).

52.    S. Sharma and M. M. Turner, J. Phys. D: Appl. Phys. **46** (285203) (2013).

53.    P. C. Boyle, A. R. Ellingboe and M. M. Turner, J. Phys. D: Appl. Phys. **37**, 697 (2004).

54.    S. Sharma and M. M. Turner, J. Phys. D: Appl. Phys. **47** (28), 285201 (2014).

55.    P. D. t. S. Sharma, *Investigation of ion and electron kinetic phenomena in capacitively coupled radio-frequency plasma sheaths: A simulation study*. (Dublin City University, Dublin City University, 2013).

56.    B. G. Heil, U. Czarnetzki, R. P. Brinkmann and T. Mussenbrock, J. Phys. D: Appl. Phys. **41**, 165202 (2008).

57.    U. Czarnetzki, J. Schulze, E. Schungel and Z. Donko, Plasma Sources Sci. Technol. **20**, 024010 (2011).

58.    B. Bruneau, T. Novikova, T. Lafleur, J. P. Booth and E. V. Johnson, Plasma Sources Sci. Technol. **23**, 065010 (2014).

59.    B. Bruneau, T. Gans, D. O'Connell, A. Greb, E. Johnson and J.-P. Booth, Phys. Rev. Lett. **114**, 125002 (2015).

60.    E. Schungel, I. Korolov, B. Bruneau, A. Derzsi, E. Johnson, D. O'Connell, T. Gans, J. P. Booth, Z. Donko and J. Schulze, J. Phys. D: Appl. Phys. **49**, 265203 (2016).

61.    X. V. Qin, Y. H. Ting and A. E. Wendt,  **19**, 065014 (2010).

62.    H. Shin, W. Zhu, L. Xu, V. M. Donnelly and D. J. Economou, Plasma Sources Sci. Technol. **20**, 055001 (2011).

63.    D. J. Economou, J. Vac. Sci. Technol. A **31**, 050823 (2013).

64.    T. Lafleur, Plasma Sources Sci. Technol. **25**, 013001 (2016).

65.    S. Sharma, S. K. Mishra, P. K. Kaw, A. Das, N. Sirse and M. M. Turner, Plasma Sources Sci. Technol. **24**, 025037 (2015).

66.    S. Sharma, S. K. Mishra, P. K. Kaw and M. M. Turner, Phys. Plasmas **24** (1), 013509 (2017).

67.    S. Sharma, N. Sirse and M. M. Turner, Plasma Sources Sci. Technol. **29** (11), 114001 (2020).

68.    S. Sharma, N. Sirse, A. Kuley and M. M. Turner, Phys. Plasmas **28** (10), 103502 (2021).

69.    S. Sharma, N. Sirse, A. Kuley and M. M. Turner, J. Phys. D: Appl. Phys. **55** (27), 275202 (2022).

70.    S. Sharma, N. Sirse, P. K. Kaw, M. M. Turner and A. R. Ellingboe, Phys. Plasmas **23**, 110701 (2016).

71.    R. Shahid, B. Kallol and C. Ken, Plasma Sources Sci. Technol. **19**, 015014 (2010).

72.    S. Wilczek, J. Trieschmann, J. Schulze, E. Schuengel, R. P. Brinkmann, A. Derzsi, I. Korolov, Z. Donko and T. Mussenbrock, Plasma Sources Sci. Technol. **24**, 024002 (2015).

73.    P. A. Miller, E. V. Barnat, G. A. Hebner, P. A. Paterson and J. P. Holland, Plasma Sources Sci. Technol. **15**, 889-899 (2006).

74.    R. R. Upadhyay, I. Sawada, P. L. G. Ventzek and L. L. Raja, J. Phys. D: Appl. Phys. **46**, 472001 (2013).

75.    S. Sharma, A. Sen, N. Sirse, M. M. Turner and A. R. Ellingboe, Phys. Plasmas **25**, 080705 (2018).

76.    S. Sharma, N. Sirse, A. Sen, J. S. Wu and M. M. Turner, J. Phys. D: Appl. Phys. **52**, 365201 (2019).

77.    S. Sharma, N. Sirse, M. M. Turner and A. R. Ellingboe, Phys. Plasmas **25**, 063501 (2018).

78.    S. Wilczek, J. Trieschmann, J. Schulze, Z. Donko, R. P. Brinkmann and T. Mussenbrock, Plasma Sources Sci. Technol. **27**, 125010 (2018).

79.    S. Sharma, N. Sirse, A. Sen, M. M. Turner and A. R. Ellingboe, Phys. Plasmas **26**, 103508 (2019).

80.    S. Sharma, N. Sirse, A. Kuley and M. M. Turner, Plasma Sources Sci. Technol. **29**, 045003 (2020).

81.    C. K. Birdsall, IEEE Trans. Plasma Sci. **19** (2), 65 (1991).

82.    Shu Zhang, Guang-Yu Sun, Arnas Volcokas, Guan-Jun Zhang and A.-B. Sun, Plasma Sources Sci. Technol. **30**, 055007 (2021).

83.    J. U. Brackbill and D. W. Forslund, Journal of Computational Physics **46**, 271-308 (1982).

84.    B. I. Cohen, A. B. Langdon and A. Friedman, Journal of Computational Physics **46**, 15-38 (1982).

85.    B. I. Cohen, A. B. Langdon and D. W. Hewett, Journal of Computational Physics **81**, 151-168 (1989).

86.    M. R. Gibbons and D. W. Hewett, Journal of Computational Physics **120**, 231 (1995).

87.    H. R. Lewis, Journal of Computational Physics **6** (1), 136-141 (1970).

88.    V. Vahedi, G. DiPeso, C. K. Birdsall, M. A. Lieberman and T. D. Rognlien, Plasma Sources Sci. Technol. **2**, 261 (1993).

89.    E. Kawamura, C. K. Birdsall and V. Vahedi, Plasma Sources Sci. Technol. **9**, 413 (2000).

90.    A. Friedman, S. E Parker, S. L. Ray and C. K. Birdsall, Journal of Computational Physics **96**, 54 (1991).

91.    Matthew R. Gibbons and D. W. Hewett, Journal of Computational Physics **120**, 231 (1995).

92.    S. Mattei, K. Nishida, M. Onai, J. Lettry, M.Q. Tran and A. Hatayama, Journal of Computational Physics **350**, 891-906 (2017).

93.    Justin Ray Angus, Anthony Link, Alex Friedman, Debojyoti Ghosh and J. D. Johnson, Journal of Computational Physics **456**, 111030 (2022).

94.   https://github.com/PrincetonUniversity/EDIPIC-2D,  (2022).

95.   A. T. Powis, W. Villafana and I. D. Kaganovich, in *International Electric Propulsion Conference* (Cambridge, MA, 2022).

96.   V. Vahedi and M. Surendra, Comput. Phys. Commun. **87**, 179 (1995).

97.   V. Vahedi and G. DiPeso, Journal of Computational Physics **131**, 149-163 (1997).

98.   A. Bruce Langdon, Bruce I Cohen and A. Friedman, Journal of Computational Physics **51** (1), 107-138 (1983).

99.   D. C. Barnes and L. Chacón, Computer Physics Communications **258**, 107560 (2021).

100.   H. Zhang, E. M. Constantinescu and B. F. Smith, SIAM Journal on Scientific Computing **44** (1), C1-C24 (2022).

101.   J. Squire, H. Qin and W. M. Tang, Phys. Plasmas **19** (8), 084501 (2012).

102.   Jianyuan Xiao, Q. Hong and L. Jian, Plasma Sci. Technol. **20** (11), 110501 (2018).

103.   A. S. Glasser and H. Qin, J. Plasma Phy. **86** (3), 835860303 (2020).

104.   C. K. Birdsall and A. B. Langdon, *Plasma physics via computer simulation*. (CRC press, 2004).

105.   M. Hayashi, Technical Report NIFS-DATA-72 (2003).

106.   S. Pancheshnyi, S. Biagi, M. C. Bordage, G. J. M. Hagelaar, W. L. Morgan, A. V. Phelps and L. C. Pitchford, Chemical Physics **398**, 148-153 (2012).

107.   S. A. Maiorov, Plasma Physics Reports **35**, 802-812 (2009).

108.   B. G. Heil, IEEE Transactions on Plasma Science **36** (4) (2008).

109.   D. Eremin, Journal of Computational Physics **452**, 110934 (2022).

110.   M. M. Turner, Phys. Plasmas **13**, 033506 (2006).

111.   S.V. Berezhnoi, I.D. Kaganovich and L. D. Tsendin, Plasma Phys. Reports **24**, 556-563 (1998).

112.   S.V. Berezhnoi, I.D. Kaganovich and L. D. Tsendin, Plasma Sources Sci. Technol. **7**, 268-281 (1998).

113.   R. D. Falgout and U. M. Yang, in *In Computational Science—ICCS 2002: International Conference Amsterdam* (Berlin, Heidelberg: Springer Berlin Heidelberg., The Netherlands, 2002), pp. 632-641.

114.   V. Vahedi and M. Surendra, Computer Physics Communications **87** (1-2), 179-198 (1995).

115.   T. Charoy, J. P. Boeuf, A. Bourdon, J. A. Carlsson, P. Chabert, B. Cuenot, D. Eremin, L. Garrigues, K. Hara, I. D. Kaganovich, A. T. Powis, A. Smolyakov, D. Sydorenko, A. Tavant, O. Vermorel and W. Villafana, Plasma Sources Science and Technology **28 (10)**, 105010 (2019).